# Growth Mechanisms and Mechanical Response of 3D Superstructured Cubic and Hexagonal $Hf_{1-x}Al_xN$ Thin Films

Marcus Lorentzon[1,*], Naoki Takata[2], Diederik Depla[3], Tianqi Zhu[4], Grzegorz Greczynski[1], Rainer Hahn[5], Anton Zubayer[1], Justinas Palisaitis[1], Helmut Riedl[5], Dasom Kim[2], Lars Hultman[1, 5], Jens Birch[1], Naureen Ghafoor[1]

[1] Thin Film Physics Division, Department of Physics, Chemistry and Biology (IFM), Linköping University, SE-58183 Linköping, Sweden

[2] Department of Materials Design Innovation Engineering, Graduate School of Engineering, Nagoya University, Nagoya, Japan

[3] Department of Solid State Sciences, Ghent University, Krijgslaan 281 (S1), 9000 Gent, Belgium

[4] Department of Mechanical, Manufacturing & Biomedical Engineering, Trinity College Dublin, The University of Dublin, College Green, Dublin 2, Ireland

[5] Christian Doppler Laboratory for Surface Engineering of high-performance Components, TU Wien, A-1060 Wien, Austria

[6] Center for Plasma and Thin Film Technologies, Ming Chi University of Technology, 84 Gungjuan Rd., Taishan Dist. New Taipei City 24301, Taiwan

*marcus.lorentzon@liu.se, +4670 - 88 71 745

# Abstract

Transition metal aluminum nitrides are a technologically important class of multifunctional ceramics, however, the HfAlN system remains largely unexplored. We investigate phase stability, nanostructure design, and mechanical behavior of $Hf_{1-x}Al_xN_y$ thin films deposited on MgO(001) substrates using ion-assisted reactive magnetron sputtering. Compared to growth temperature and ion assistance, backscattered Ar neutrals are shown to have a dominant influence on the film structure. The Al-rich ($x > 0.41$) films form a nanocrystalline morphology consisting of Hf- and Al-rich nanodomains in a wurtzite-hexagonal(h) 0001 fiber-texture exhibiting about 22 GPa hardness, considerably higher than that of a binary AlN. For low Al contents, $x < 0.30$, surface-driven spinodal decomposition by energetic Ar neutrals during deposition in combination with quenching of sub-surface diffusion results in an unusual – and unique for nitrides - three-dimensional checkerboard superstructure of AlN- and HfN-rich nanodomains in the single-crystal rocksalt-cubic (c) phase. Lattice-resolved scanning transmission electron microscopy complemented with x-ray and electron diffraction reveals that the superstructure periodicity extends along <100> directions and the size increases linearly from 9 to 13 Å with rising Al content. Consequently, the nanoindentation hardness increases sharply from 26 GPa for $HfN_y$, to ~38 GPa for c-$Hf_{1-x}Al_xN_y$, due to dislocation pinning at the superstructure strain fields. Micropillar compression of c-$Hf_{0.93}Al_{0.07}N_{1.15}$ shows a considerably higher yield stress compared to $HfN_y$ and controlled brittle fracture occurs via {110}<011> slip systems, attributed to superstructure inhibited dislocation motion. In contrast, nanocrystalline h-$Hf_{0.59}Al_{0.41}N_{1.23}$ exhibits a high yield stress and limited plasticity before strain burst failure.

# Graphical Abstract

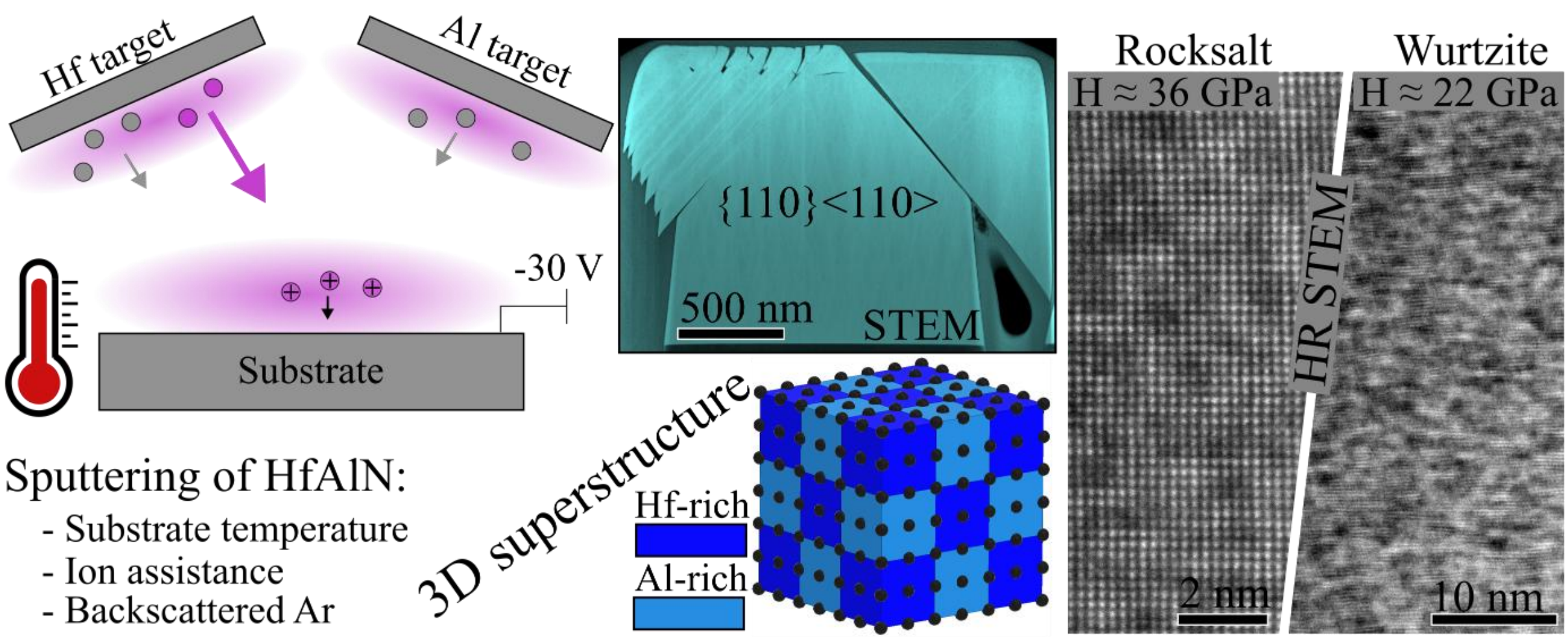

## 1. Introduction

Transition metal aluminum nitrides (TM-Al-N) are an industrially important class of ceramic materials used as hard and wear protective coatings with multiple functional properties. The addition of AlN into binary cubic(c) rocksalt c-TM-Ns is a common method to boost the properties such as hardness, elastic modulus, thermal stability, oxidation resistance, and electrical conductivity [1], [2], [3], [4], [5]. In addition, the good thermal stability and density of TM-N allows application as diffusion barrier in microelectronics [5], [6]. A high Al content improves the oxidation resistance of cubic phase coatings, where the amount of AlN possible to incorporate into the rocksalt lattice depends on choice of TM and growth conditions. Above a certain composition, the hexagonal(h) wurtzite structure (h-AlN) becomes energetically more favorable, causing phase transformation and concurrently a change in coating properties, in particular loss of hardness. The structure and properties are well-studied for the TiAlN material system [7], known to form metastable alloys of rocksalt solid solution in excess of 60% Al on the metal sublattice. The alloys may undergo spinodal decomposition either during film deposition or at post-deposition high-temperature annealing, such as friction heating during coating operation on cutting tools. Growth-initiated decomposition results in elongated nanosized c-TiN and c-AlN domains along the substrate normal [8], while temperature-initiated (annealing) spinodal decomposition results in three-dimensional domains of the same components, often with some retained c-TiAlN matrix [7], [9]. Both phenomena results in an age-hardening effect, i.e., the hardness of the coating increases with temperature, due to the evolving nanosized chemical modulation of Al-rich and Ti-rich domains [10] where strain fields between the isostructural coherent domains hinder dislocation motion.

The driving force for decomposition of metastable c-TM-Al-N solid solutions is the positive enthalpy of mixing of the pseudo-binary TM-N and Al-N phases [11], [12], [13]. In comparison to TiAlN, other group IV alloys, i.e., ZrAlN and HfAlN have nearly twice as high enthalpy of mixing, which affects the decomposition route and phase transformation, and therefore also the associated properties. For HfAlN and ZrAlN the lattice mismatch between c-TM-N and c-AlN is large, 11.9 % and 13.2 % respectively, compared to 4.9 % for c-TiN and c-AlN, calculated using ICDD PDF cards for HfN 00-033-0592, ZrN 00-035-0753, TiN 00-038-1420 and c-AlN 00-046-1200. The strain energy that develops between coherent domains may therefore counteract the chemical driving force for decomposition, in particular for HfAlN and ZrAlN, eventually balancing the total driving force to near zero [11]. Sheng et al. theoretically predicted that for ZrAlN, the strain energy from coherent domains of c-ZrN and c-AlN is sufficiently large that instead of isostructural decomposition the material decomposes directly into c-ZrN and h-AlN [13], experimentally demonstrated by Ghafoor et al. [14].

HfAlN, compared to TiAlN and ZrAlN, has been much less studied with only a few reports in literature, e.g. [15], [16], [17]. The theoretical limit for metastable cubic phase $Hf_{1-x}Al_xN$ is predicted to $x \approx 0.45$ [11]. However, up to $x \approx 0.51$ has been synthesized using ion-assisted DC magnetron sputtering, and vague indications of the material undergoing surface-initiated spinodal decomposition has been shown [15]. HfN’s exceptional thermal stability and high melting point of ~3300 °C [18] make its alloys a strong candidate for ultra-high-temperature applications [19]. Beside protective coating applications, the tunable properties make it an attractive material system for plasmonic optical applications [20], [21], renewable energy [22], and as diffusion barriers in electronics [23], [24].

Here, we study the structure of $Hf_{1-x}Al_xN_y$ (x = 0 - 0.7, y > 1) films deposited on lattice matched MgO(001) substrates using ion-assisted reactive magnetron sputtering in a mixed Ar and $N_2$ atmosphere. The effect of substrate temperature, ion assistance and backscattered Ar neutrals on the resulting film structure are investigated, where in particular abundant backscattered high energy Ar neutrals have a large effect on the film formation [25], when sputtering heavy elements such as Hf. The threshold composition for the cubic to hexagonal transition is revisited under these growth conditions, which promote high adatom diffusivity for higher crystalline quality [26], [27], with the intention of promoting superior mechanical properties.

The article first presents the film composition, bonding, and phase analysis followed by a detailed structural comparison of the cubic and hexagonal phases as influenced by growth conditions. It continues with an in-depth analysis of how the structure impacts mechanical properties including hardness, strength and failure mechanisms. Particular attention is given to the compositional dependence of phase formation and resulting nanostructures. The results reveal the emergence of a self-organized, three-dimensional superstructure in the cubic phase (c- $Hf_{1-x}Al_xN_y$) for Al contents below $x \approx 0.3$, characterized by a chemical modulation period below 13 Å that scales nearly linearly with Al concentration. This phase exhibits a stable high hardness, a high yield strength and controlled brittle fracture. Upon exceeding the cubic solubility limit at $x > 0.41$, the system transitions to a fiber-textured nanocrystalline hexagonal phase (h-$Hf_{1-x}Al_xN_y$), which exhibits a lower hardness, high yield strength and some plasticity before uncontrolled brittle fracture and catastrophic failure.

# 2. Experimental Details

## 2.1. Film growth

The $Hf_{1-x}Al_xN_y$ films were grown in a high vacuum deposition chamber on MgO(001) and Si(001) substrates using two magnetically coupled ϕ = 75 mm (3") unbalanced type II magnetrons. Elemental targets of Hf (99.9% except for a few percent Zr) and Al (99.99 %) were co-sputtered in an Ar + $N_2$ atmosphere at 0.6 Pa (4.5 mTorr) with a $N_2$ partial pressure of 0.067 Pa (0.5 mTorr), achieved by an Ar and $N_2$ flow rates of 43 sccm and 8 sccm, respectively. The base pressure was better than $6.4*10^{-4}$ Pa ($4.8*10^{-6}$ Torr) at the deposition temperature of 800 °C. Details of the deposition system can be found elsewhere [28], [29]. The MgO substrates were prepared according to the method described in Ref [30]: cleaned by sonication in a detergent solution (Hellmanex III, 2 vol.%), for 5 min, rinsed in de-ionized water followed by 10 min sonication in acetone and ethanol. Finally, the substrates were blow dried with $N_2$ just before inserting into the deposition chamber. An annealing treatment was performed for 1 h at 900 °C prior to deposition.

A high flux of low energy ion assistance was applied by coupling to the target magnetic field, thus extending the plasma down to the substrate table using a stationary electromagnetic coil around the rotating (16 rpm) substrate table. A -30 V substrate bias was used to attract ions to bombard the growing film which enhances the adatom mobility. A sharp Mo-pin was used to make electrical contact to the conducting film since MgO is insulating.

By varying the applied powers to the targets, two composition series were deposited. Six films were deposited for 2 h in Series A, where the sum of the magnetron powers was held constant at 250W ($P_{Hf}$ + $P_{Al}$ = 250 W), while the power ratios $P_{Hf}/P_{Al}$ varied: 250/0, 200/50, 175/75, 150/100, 125/125, and 75/175. For this series, the magnetic field was coupled with the Al-target, except for the HfN film where the coil current was reversed to couple with the Hf-target. The resulting film thicknesses were between 2.5 μm to 1.2 μm, where the growth rate decreased with increasing Al powers. In the second, Series B, six films were deposited for 2.5 h, focused solely on the cubic phase HfAlN while attempting to hold the ion assistance parameters as constant as possible. $P_{Hf}$ was held constant at 125 W with the magnetic field coupled to the Hf magnetron. $P_{Al}$ was varied between 0 W and 80 W (0, 20, 30, 40, 60, 80 W). The resulting film thicknesses varied between 0.93 and 1.3 μm, increasing with increasing Al power. To investigate the effects of substrate temperature, one film from Series B was selected, Al-power of 60 W, and redeposited at 300 °C, 600 °C and 900 °C. All other deposition parameters were kept constant. Additionally, all films in both series were re-deposited with a thickness of about 200 nm on Si(001) substrates (with a native surface oxide) at otherwise identical conditions to accurately determine the chemical composition using ion-beam analysis techniques. The Si substrates were cleaned before deposition by sonication for 5 min in acetone and isopropanol.

## 2.2. Plasma analysis

Langmuir and area plasma probes were used to measure the plasma potential, $V_p$, floating potential, $V_f$, and ion saturation current, $I_{sat}$, at the substrate position. The Langmuir probe was a W wire with ϕ = 0.5 mm, sticking out 4 mm from an insulated ceramic tube. $V_p$ was determined from the measured I-V curve by fitting the linear segments in the log plot and taking the value where the lines cross. $V_f$ was determined at the point of zero current, fairly constant between -26 V and -32 V, i.e. similar to the applied bias of -30 V. The ion energy was calculated as $E_{ion} = q|V_p - V_{bias}|$, where q was taken as the elemental charge. The area probe, made from a 1 $cm^2$ stainless steel disc mounted flush but electrically insulated to a stainless-steel holder which was kept at the same potential, was used to measure the ion current as function of the applied voltage. $I_{probe}$ was taken at -100 V and the ion flux, $J_{ion}$, was calculated using equation *2.1*, where γ = 10% is the secondary electron emission.

$$J_{Ion} = I_{probe}\frac{(1-\gamma)}{q} \quad 2.1$$

## 2.3. Chemical analysis

A combination of time-of-flight elastic recoil detection analysis (ToF-ERDA) and Rutherford backscatter spectrometry (RBS) techniques were used for chemical analysis of the representative 200 nm thick films at the tandem accelerator at Ångström laboratory, Uppsala University. ToF-ERDA was used to quantify the concentration of light elements in the samples such as O, N, and C. A beam of 36 MeV $^{127}I^{8+}$-ions was incident on the sample at an angle of 67.5° with respect to the surface-normal and detectors were placed at a 45° angle with respect to the incident beam in a forward scattering geometry. The ToF-detector was calibrated using standard samples (Au, TiN, CaF, SiC, and $Al_2O_3$), and energy detector efficiency files were used. The mass-resolved histograms (ToF vs Energy) was analyzed using the Potku code [31].

For RBS measurements, a 2 MeV $^4He^{2+}$ probe was incident on the sample at 5° with respect to the sample normal and the detector placed at a 170° backscattering angle. The sample was continuously rocked back and forth in small random motions of 2° around the center position to avoid channeling effects. SimNRA [32] (version 7.03) was used to simulate and fit the experimental data and reference samples (Au, Cu, TiN, SiC, and C) were used to calibrate the energy scale. To account for the small effects of multiple scattering causing background noise, trace amounts (<0.3 at.%) of a heavy element (U) was used in the simulations. Together with ToF-ERDA results, the accuracy of the composition is estimated to be better than 0.02 for the Hf-to-Al ratio, and 0.03 for the metal-to-nitrogen ratio.

High-resolution X-ray photoelectron-spectroscopy (XPS) spectra of Hf 4f, Al 2p, Al 2s, N 1s, and O 1s, were acquired in a Kratos AXIS Ultra spectrometer at a base pressure of $2.0*10^{-9}$ mbar. A 0.3 × 0.7 $mm^2$ sample area was analyzed using monochromatic Al Kα radiation with the excitation source operated at 150W (15 keV, 10 mA). The electron emission angle was normal to the sample surface. Gentle 0.5 keV $Ar^+$ etching at a shallow ion incidence angle of 20° from the surface plane was applied for 720 s, to remove surface oxides while minimizing the sputter damage [33]. Spectrometer binding energy (BE) scale was calibrated according to the ISO 15472:2010 standard. Charge referencing was done using the Fermi edge of the conducting samples (see supplementary information section 1) to avoid problems inherent to the adventitious carbon method [34]. Spectra from samples with higher Al content (x > 0.41) that do not exhibit a pronounced Fermi edge are aligned against the Al 2p peak, which does not differ substantially between cubic or wurtzite phase, similar to the ZrAlN system [14], [35].

## 2.4. Structural analysis

X-ray diffraction (XRD) θ-2θ scans were carried out in a Panalytical X'Pert Pro in Bragg-Brentano geometry with 1/2° slits and monochromatic Cu Kα radiation by using a Ni-filter. ω-scan rocking curves were acquired in a Panalytical Empyrion using a hybrid monochromator and 3-axis analyzer giving a full width at half maximum (FWHM), $\Gamma_\omega$, of 0.013° for 002 reflection of MgO(001) substrate peak. X-ray reflectivity (XRR) was performed on the thinner (~200 nm) samples on Si substrate, to extract density information, using the known composition from RBS. The measurement was performed in a Rigaku SmartLab diffractometer with Cu kα radiation from a parallel beam by a double crystal monochromator (CBO) and a 5 mm mask to ensure a good footprint on the sample. The XRR curves between 2θ = 0° - 3° were fitted using GenX3 [36].

TEM samples were prepared with the conventional sandwich method by gluing two pieces of the sample together in a TEM grid with the film facing each other, followed by mechanical polishing to ~50 μm thickness and thinning to electron transparency by $Ar^+$ ion milling with 5 keV at 5°. The final step of ion polishing using 1.5 keV for 25 min was used to remove amorphized surface damage. Selected area electron diffraction (SAED) patterns were acquired in a FEI Tecnai G2 instrument operated at 200 kV using a 40 μm diameter aperture. Lattice resolved high angle annular dark field scanning transmission electron microscopy (HAADF STEM) micrographs were obtained in a double Cs corrected $Titan^3$ 60-300 microscope, operated at 300 kV. Samples were aligned to the MgO(001) zone axis for imaging or diffraction.

Films were analyzed by wide angle x-ray scattering (WAXS) at P07 High Energy Materials Science beam line at Petra III, DESY in Hamburg. The full experimental setup is presented elsewhere [37]. A monochromatic x-ray beam with energy of 73.8 keV and beam size of ~1×1 $mm^2$ was incident on the sample in transmission geometry with the x-ray beam perpendicular to the sample normal. The sample was placed on a Mo sample stage inside a vacuum chamber equipped with two x-ray transparent sapphire windows. A Perkin Elmer area detector collected the diffracted intensity. The data was calibrated by measuring the Debye-Scherrer rings of a standard polycrystalline $LaB_6$ sample, whereafter the experimental setup could be calibrated for sample-detector distance, detector tilt and x-y position. Image calibration and analysis was performed using the python library pyFAI [38].

## 2.5. Hardness and Micropillar Compression Testing

Hardness was measured by nanoindentation in a Triboindenter, Hysitron TI950, equipped with a Berkovich diamond tip. The area function was calibrated by indents in a fused quartz standard. The load-displacement curves were analyzed by the Oliver and Pharr method [39], and 25 indents were made for each sample.

Cylindrical micropillars for compression testing were milled from the surface of c-HfAlN sample with a diameter of ~1 μm (with ~5.8° taper) and an aspect ratio of ~2:1 using a JEOL JEM-9320 focused ion beam (FIB) system operated at 30 kV. Similarly, micropillars from h-HfAlN were fabricated using a ThermoFisher Scios 2 FIB and scanning electron microscope (SEM) workstation. Each pillar had a diameter of approximately 400 nm and an aspect ratio of 3:1. The fabrication process employed a $Ga^+$ ion beam, with a probe current of 7 nA for coarse milling and subsequently reduced to 50 pA for final polishing. Particular care was taken to terminate the milling precisely at the coating–substrate interface to ensure consistent boundary conditions. The final taper angle was below 2° for all pillars.

Compression tests were carried out in a load-controlled nanoindenter SHIMADZU DUH-211S equipped with a 20 μm flat-punch diamond tip at ambient temperature. An initial strain rate of all micropillars was controlled to approximately $10^{-2}$ $s^{-1}$. The compressed pillars were imaged at a ~50° tilt angle by a Zeiss Sigma 300 scanning electron microscope operated at 3 kV acceleration voltage and SE2 detector. An electron transparent lamella of a compressed micropillar was prepared by conventional lift-out method using Carl Zeiss Crossbeam 1540 EsB FIB-SEM workstation. Before thinning and lift out, an electron beam-induced Pt layer was deposited to provide protection from gallium implantation. For final thinning, a low-energy milling at 2 kV was performed to minimize Ga-induced damage.

# 3. Results and Discussion

## 3.1. Elemental Analysis

The elemental composition of the two series, see Table 1, was determined by combining RBS and ToF-ERDA results obtained from 200 nm thick films on Si(001) substates; using ERDA as input to RBS data analysis and simulations. Besides Hf, Al and N, the only contaminants detected were Zr and Ar. The Zr content scales with the amount of Hf in the film, as Zr is a common contaminant in Hf targets, approximately at a level of 2 to 3 at.%. ToF-ERDA revealed a thin surface oxide layer, formed on exposure to atmosphere. The final compositions are presented as $Hf_{1-x}Al_xN_y$, where x is the Al atomic fraction on the metal sublattice and y is the N fraction with respect to the metals, i.e., a perfectly stoichiometric film with 50% Al on the metal sublattice would be written as: $Hf_{0.5}Al_{0.5}N_1$.

All films were over-stoichiometric with respect to nitrogen, with nitrogen content roughly scaling with Hf concentration. This is likely due to the high affinity between Hf and N combined with a relatively high $N_2$ partial pressure used in the experiments. A substantial amount of trapped Ar, 1.5 ± 1 at.%, is attributed to buried Ar atoms from the growth. The much higher Ar contamination found in $Hf_{0.59}Al_{0.41}N_{1.23}$ and $Hf_{0.29}Al_{0.71}N_{1.04}$, is explained by an enhanced entrapment in a distorted nanocrystalline structure, as shown below.

*Table 1: Film composition determined by ToF-ERDA and RBS. Mass density was obtained by XRR simulations using the chemical composition of the films. The crystal quality was evaluated using the FWHM of XRD rocking curves , $\Gamma_\omega^{002}$. The ion energy and ion-to-atom flux ratio were determined from plasma probe analysis combined with results from RBS simulations.*

| | ToF-ERDA & RBS | | | | | RBS & XRR | XRD | Plasma | Plasma & RBS |
|---|---|---|---|---|---|---|---|---|---|
| **Sample** | **Hf [at.%]** | **Al [at.%]** | **N [at.%]** | **Zr [at.%]** | **Ar [at.%]** | **$\rho_{mass}$ [g/cm$^3$]** | **$\Gamma_\omega^{002}$ [°]** | **$E_{ion}$ [eV]** | **$J_{ion}$ / $J_{atom}$** |
| **Serie A: Constant total magnetron power** | | | | | | | | | |
| $HfN_{1.22}$ | 44.1 | 0.0 | 53.9 | 1.2 | 0.8 | 12.55 | 0.663 | 24.4 | 7.7 |
| $Hf_{0.93}Al_{0.07}N_{1.15}$ | 42.4 | 3.0 | 52.1 | 1.2 | 1.3 | 11.70 | 0.607 | 26.1 | 4.6 |
| $Hf_{0.84}Al_{0.16}N_{1.13}$ | 38.8 | 7.2 | 52.1 | 1.0 | 0.9 | 10.51 | 0.552 | 25.9 | 6.9 |
| $Hf_{0.71}Al_{0.29}N_{1.15}$ | 32.7 | 13.1 | 52.6 | 0.8 | 1.0 | 9.04 | 0.944 | 26.0 | 9.9 |
| $Hf_{0.59}Al_{0.41}N_{1.23}$ | 25.8 | 17.6 | 53.3 | 0.7 | 2.7 | 7.78 | - | 25.5 | 12.9 |
| $Hf_{0.29}Al_{0.71}N_{1.04}$ | 14.0 | 33.9 | 49.7 | 0.3 | 2.0 | 5.71 | - | 25.0 | 21.5 |
| **Serie B: Constant ion flux** | | | | | | | | | |
| $HfN_{1.33}$ | 42.0 | 0.0 | 56.0 | 1.5 | 0.5 | 12.52 | 0.584 | 21.5 | 14 |
| $Hf_{0.94}Al_{0.06}N_{1.17}$ | 42.0 | 2.7 | 52.5 | 1.4 | 1.4 | 11.57 | 0.625 | 20.6 | 14.6 |
| $Hf_{0.90}Al_{0.10}N_{1.12}$ | 41.3 | 4.6 | 51.5 | 1.2 | 1.4 | 10.88 | 0.614 | 20.1 | 15.0 |
| $Hf_{0.85}Al_{0.15}N_{1.11}$ | 39.1 | 6.9 | 51.1 | 1.3 | 1.7 | 10.32 | 0.610 | 20.1 | 14.2 |
| $Hf_{0.76}Al_{0.24}N_{1.15}$ | 34.1 | 11.0 | 52.1 | 1.0 | 1.8 | 9.04 | 0.652 | 20.0 | 13.0 |
| $Hf_{0.70}Al_{0.30}N_{1.09}$ | 32.5 | 14.1 | 50.9 | 1.0 | 1.6 | 9.07 | 0.717 | 18.0 | 13.2 |

The mass density of the films, see Table 1, was obtained by X-ray reflectivity simulations of the ~200 nm thick films on Si(001), using the chemical composition as input, see supplementary information section 2. The excess nitrogen in $HfN_{1.22}$ and $HfN_{1.33}$ films is largely responsible for the lower density of 12.55 g/cm$^3$ and 12.52 g/cm$^3$ respectively, due to a large number of Hf-vacancies [40], compared to the stoichiometric HfN bulk density of 13.8 g/cm$^3$. Moreover, Zr contamination lowers the density, while the Ar most likely sits in interstitial positions for an increased density. We have not observed any evidence for Ar bubbles [41] or Ar-enriched clusters [42]. In Al containing films the decreasing density with increasing Al content is expected, yet the density is smaller than expected which indicates that a large number of metal vacancies and N interstitials are also present in Al containing films.

## 3.2. Phase analysis

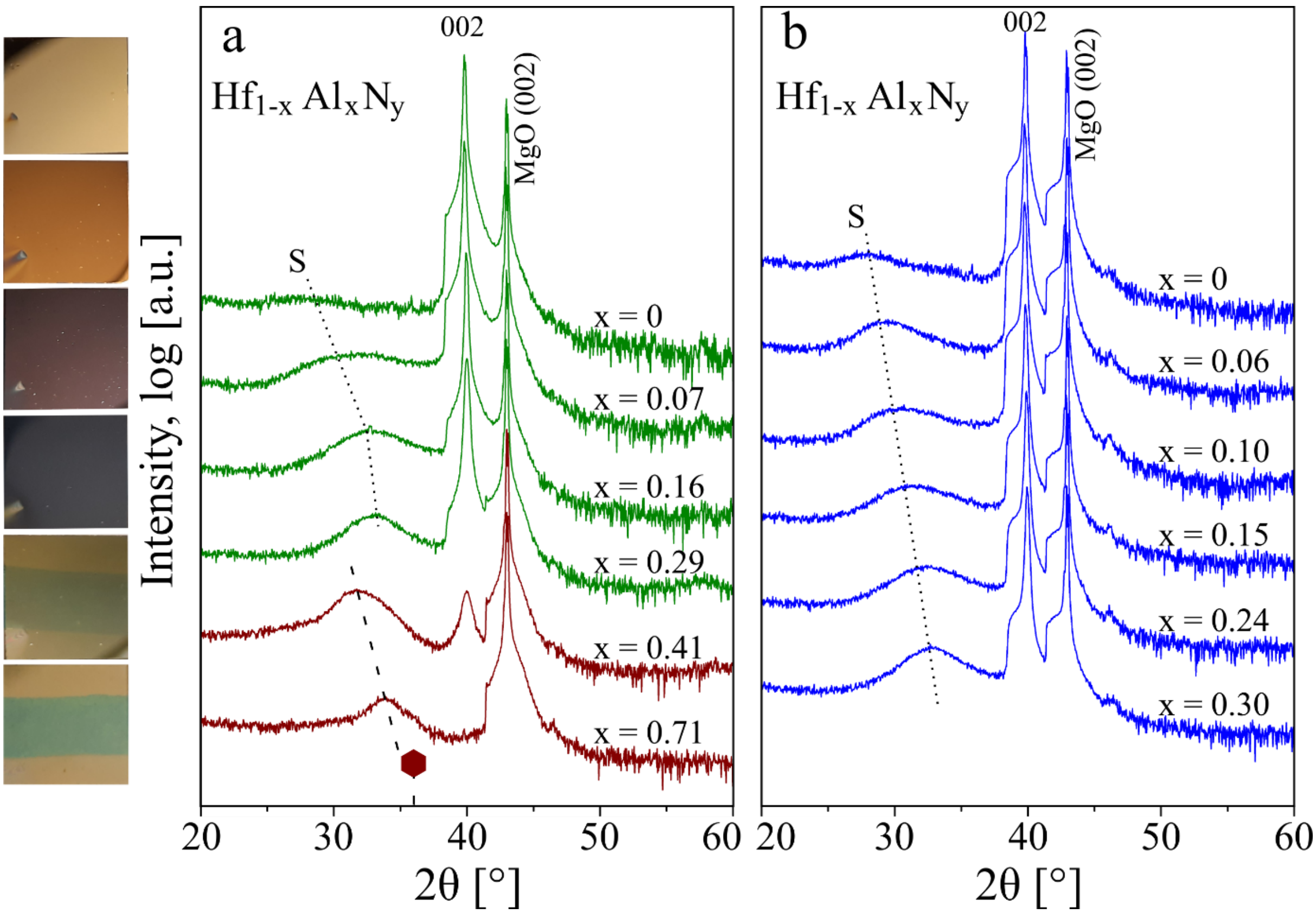

*Figure 1: Selected range of XRD θ–2θ scans from samples in a) Series A: constant total magnetron power, and b) Series B: constant ion flux. The green and blue curves show cubic rocksalt HfAlN in the two series, respectively, while red curves (x > 0.41) show wurtzite HfAlN in Series A. The red hexagon in a) marks the position of the expected AlN(0002) peak at 36.041°. Broad humps are marked with "S" in the cubic phase films. The pictures to the left show the visual appearance of the samples in a), where the cubic phase is opaque, and the two bottom pictures of wurtzite samples are transparent, shown by a blue line underneath the samples. The diffractograms for $HfN_y$ in a) and b) were reproduced with permission [40].*

XRD θ-2θ scans in Figure 1a-b (green and blue curves, respectively) show films of high-quality epitaxial rocksalt c-$Hf_{1-x}Al_xN_y$ for x < 0.3, as evidenced by the intense 002 film peaks exhibiting well-resolved $K\alpha_1$ and $K\alpha_2$ reflections, similar to those of the MgO(002) substrate. A slight peak shift towards higher 2θ angles as a function of Al-concentration shows a (partial) formation of a solid solution, with Δ2θ = 0.144° between $HfN_{1.33}$ and $Hf_{0.70}Al_{0.30}N_{1.09}$. Furthermore, the crystalline quality improves with small additions of Al, shown by the reduced FWHM, $\Gamma_\omega^{002}$, of the 002 peak in rocking curves, see Table 1. An Al fraction of about 0.15 gives the best crystal quality (excluding $HfN_{1.33}$ in Series B). This indicates that metal-vacancy point defects are being (partially) filled up by Al-atoms, thus reducing the number of voids, resulting in a lower strain and higher crystal quality. At higher Al content, x = 0.24 to 0.30, the crystal quality decreases, evident from larger values of $\Gamma_\omega^{002}$, due to induced strain fields as the ionic radii of Al is smaller than Hf [43].

For all the cubic phase films, a low intensity broad peak, or "hump", is observed at low 2θ angles. The peak position gradually shifts closer to the 002 peak with increasing Al, most clearly seen in Figure 1b. A similar hump with corresponding peak shift is also measured for the symmetric 004 peak (see supplementary information section 3). The position of the humps does not agree with any known Hf-N phase. Instead, they are attributed to satellite peaks from a compositionally modulated three-dimensional superstructure which is described in detail in Section 3.4 based on STEM and SAED analysis.

Phase transformation from cubic rocksalt to hexagonal wurtzite crystal structure is first detected for the h-$Hf_{0.59}Al_{0.41}N_{1.23}$ film, i.e. x > 0.41, with diminishing cubic 002 peak and appearance of a broad peak attributed to wurtzite HfAlN 0002, see Figure 1a (red curves). The wurtzite peak shift by Δ2θ = 4.3° towards smaller angles from the nominal position of AlN (2θ = 36.041° [PDF card 00-025-1133]) is caused by incorporation of large Hf atoms in the hcp lattice. The peak width and low intensity indicate a distorted nanocrystalline morphology of the 0002 planes. With increasing Al in h- $Hf_{0.29}Al_{0.71}N_{1.04}$, a smaller 0002 peak shift of Δ2θ=2.1° is observed while no cubic peaks are observed. DFT calculations predict that the cubic phase is energetically favored up to x = ~0.45 in a random solid solution of $Hf_{1-x}Al_xN$ [11]. For x > 0.45, the hexagonal phase or a phase mixture is instead favored. DFT calculations thus suggest that cubic films can be obtained at a higher Al concentration than our experimental observations, which show a limit closer to x = 0.30. The difference to our experiments is likely due to the dynamics of film growth, where the sputtering technique operates far from thermodynamic equilibrium.

## 3.3. Chemical bonding analysis by XPS

The Hf 4f, N 1s, and Al 2p XPS core level spectra recorded from $Hf_{1-x}Al_xN$ films in Series A are shown in Figure 2 as a function of the Al-content. For c-HfAlN films, x ≤ 0.16 (green curves), the Hf $4f_{5/2}$ and $4f_{7/2}$ spin-split peaks are at the binding energy (BE) of 17.1 eV and 15.4 eV, respectively. These values are 0.2 - 0.3 eV lower compared to an as-deposited surface, for which Hf $4f_{7/2}$ peak was found at 15.6 - 15.7 eV [44], [45]. The lower values are caused by the destructive effect of the $Ar^+$ etching, which in this case leads to the formation of a N-deficient surface layer [44] and gives rise to an extra doublet (arrow in Figure 2) in the Hf 4f spectrum. Due to the N-deficient layers' more metallic character, peak shifts to lower BE with respect to the signal from the as-deposited nitride that is still present below the $Ar^+$-affected surface region [31]. The N 1s peak from $HfN_{1.22}$, x = 0, is at 397.80 eV, i.e., shifted to higher BE by ~0.6 eV with respect to pristine HfN, due to N understoichiometry resulting from $Ar^+$ etching [44], [46], which also accounts for  shoulders on the low-BE side of the primary N 1s peak [46].

The Hf 4f spectrum from c-$Hf_{0.71}Al_{0.29}N_{1.15}$, x = 0.29, clearly exhibits two doublets, where in addition to the original signal, a new doublet appears, shifted by ~1.1 eV towards higher BE. Concurrently, the N 1s signal exhibits a second component shifted by ~0.8 eV towards lower BE, thus revealing the formation of a different phase from the cubic. By examining the Hf 4f and N 1s spectra for h-$Hf_{1-x}Al_xN$, x > 0.41, where the new doublet increases in intensity while the doublet from the cubic phase diminishes, we attribute the new doublet to the binding energy of wurtzite phase h-$Hf_{1-x}Al_xN$. No large changes are observed in the Al 2p signal, besides the expected increase in the signal intensity with increased Al concentration. The BE of 74.2 eV observed for the Al 2p peak agrees well with the values published for AlN [44]. This clearly shows a distinct difference in the chemical environment experienced by the Hf and N atoms in the cubic and hexagonal crystal structures, while the Al atoms are rather unaffected. The shift in BE suggests a stronger charge transfer from Hf to N in the wurtzite structure, which may be explained by the different bond lengths in wurtzite, compared to cubic rocksalt: Each atom A has four nearest neighbors of atom B, similar to cubic rocksalt, but one of the neighbors has a larger bond length compared to three more closely spaced neighbors. Similar effects have been shown in ZrAlN, where the Zr 3d peaks shifted with ~1.7 eV to higher BE in wurtzite compared to cubic structure [35].

The XPS and XRD (Figure 1), show few contradictions in phase analysis. According to XRD, supported by TEM and SAED analysis below, $Hf_{0.71}Al_{0.29}N_{1.15}$ are composed of a single cubic phase while Hf 4f and N 1s spectra in XPS (cf. orange curves in Figure 2) reveal a phase mixture. In fact, also the Hf 4f spectrum from $Hf_{0.84}Al_{0.16}N_{1.13}$ shows a clear deviation from the expected 3:4 area ratio for the 4f doublet, which may indicate that a minor fraction of the hexagonal phase is present already at this low Al content. These discrepancies are explained by sputter-induced phase transformations occurring at the surface region exposed to the $Ar^+$ etching prior to XPS analyses. Even though the incident ion energy is intentionally low at 500 eV to limit sputter damage, this extra energy transfer leads to atomic re-arrangement to form supersaturated domains, with concentrations above or below the critical level for transition to either the wurtzite or rocksalt phase. Thus, in high Al-containing $Hf_{1-x}Al_xN$, both cubic and wurtzite phase co-exist in the sputter-affected surface.

These results demonstrate that the interpretation of XPS spectra acquired from metastable materials requires a high degree of caution and often the input from complementary techniques is necessary to avoid false conclusions unless data are obtained from a pristine surface, retained by utilizing the technique of a thin capping layer [44], [47], or by *in-situ* XPS.

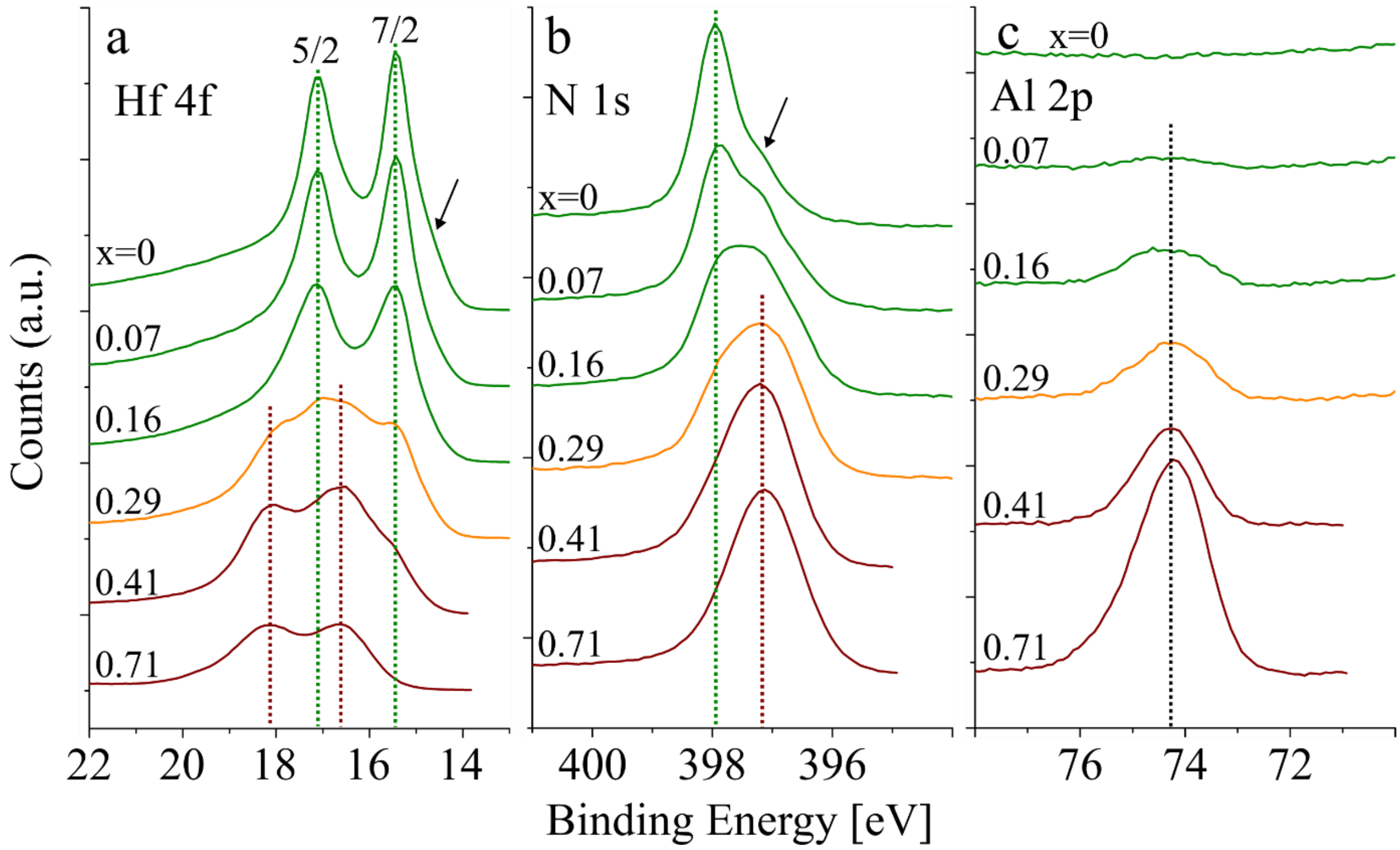

*Figure 2: Core level a) Hf 4f, b) N 1s, and c) Al 2p high resolution XPS spectra of the HfAlN films in Series A. Green curves correspond to spectra obtained from films dominated by cubic phase while red curves to films dominated by wurtzite phase. The orange curves, $Hf_{0.71}Al_{0.29}N_{1.15}$, shows a clear mixture of both cubic and wurtzite doublets. The dotted green lines mark the BE from cubic phase while the dotted red lines mark the BE from wurtzite phase. The arrows indicate signal from understoichiometric top surface layer due to $Ar^{+}$-etching.*

## 3.4. Superstructure formation in c-$Hf_{1-x}Al_xN_y$

Lattice-resolved HAADF STEM, SAED, and synchrotron WAXS were used to investigate the structural details of the as-deposited c-HfAlN films. The HAADF STEM micrographs from c-$Hf_{1-x}Al_xN$, x = 0.07, 0.16, and 0.30 in Figure 3a-c shows single crystalline cubic lattices for all three compositions. Moreover, nanometer-sized areas of bright and dark contrast superimposed onto the lattice are attributed to three-dimensional cubic Hf-rich and Al-rich domains, respectively, a hallmark of spinodal decomposition [8], [15], [48]. The contrast between the domains is more pronounced at higher Al content as expected from a larger mass difference.

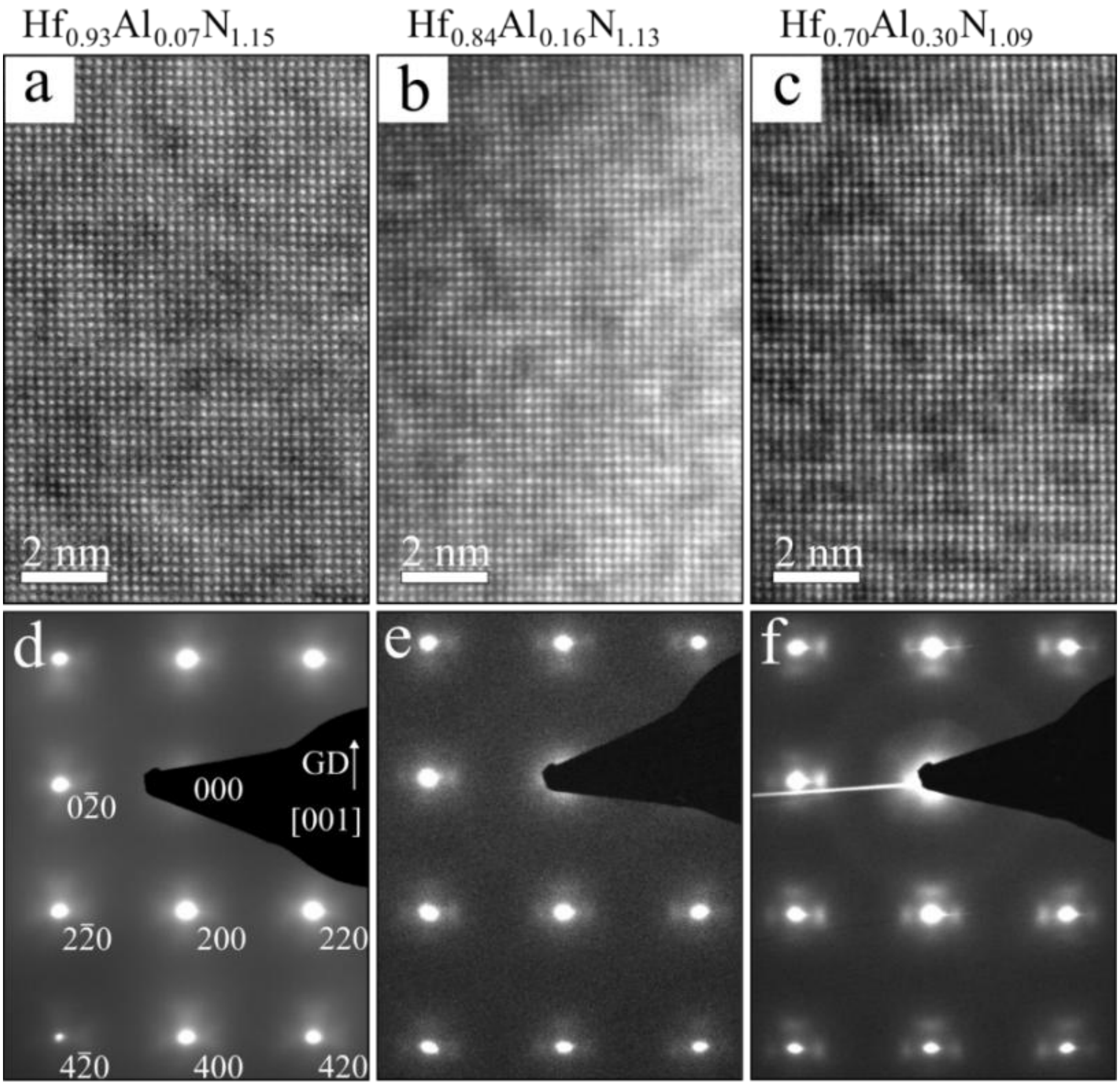

*Figure 3: a)-c) Cross-sectional HR HAADF STEM micrographs along the [001] zone axis of cubic-phase $Hf_{0.93}Al_{0.07}N_{1.15}$, $Hf_{0.84}Al_{0.16}N_{1.13}$ and $Hf_{0.70}Al_{0.30}N_{1.09}$ films. d)-f) corresponding SAED patterns, showing superstructure satellite lobes around each lattice spot. The indexing in d) applies to e)-f) as well.*

The Hf- and Al-rich domains self-organize into a compositionally modulated superstructure in a checkerboard-like pattern along <001> directions, best observed in the lower parts of Figure 3c where alternating areas of bright and dark contrast can be seen. This pattern introduces a secondary periodicity to the material, beyond the lattice itself. As a consequence, satellite peaks appear in reciprocal space, clearly observed in SAED in Figure 3d-f, where satellites surround each of the main single-crystal diffraction spots, indexed in Figure 3d. The satellites are equidistant from the main spot and oriented along <001> directions, which confirms the checkerboard orientation in Figure 3a-c. The large peak width of the satellites indicates a relatively large size distribution of the Hf-rich and Al-rich nitride domains. Note that, since the TEM lamella was taken from a cross-section of the film, the satellites are obtained in both lateral (in-plane) and transverse (out-of-plane) directions. SAED from a plan-view TEM lamella (supplementary information section 4) confirms the three-dimensional nature of the checkerboard superstructure.

Observations by STEM and SAED provide local information from thin, electron-transparent lamellae; however, the sample preparation process can, in the worst case, induce structural changes, potentially resulting in features that are not fully representative of the as-deposited film. To rule out such artifacts and confirm that the 3D superstructure homogeneously exists in the as-deposited film, we performed wide-angle synchrotron scattering on $Hf_{0.84}Al_{0.16}N_{1.13}$, shown in Figure 4a. The high brilliance of the synchrotron enables diffraction in transmission geometry from the entire 10 × 10 mm$^2$ sample. The clearly distinguishable satellites surrounding the main peaks confirm the checkerboard superstructure. Thus, we conclude that the electron diffraction and STEM micrographs (cf. Figure 3e) are indeed representative of the overall film structure. HAADF STEM images at slightly lower magnifications provide easy identification of the superstructure from the speckled contrast as shown in Figure 4b. Broad satellites surrounding the main spots are also present in the fast-Fourier transform (FFT) of the image, shown in the inset.

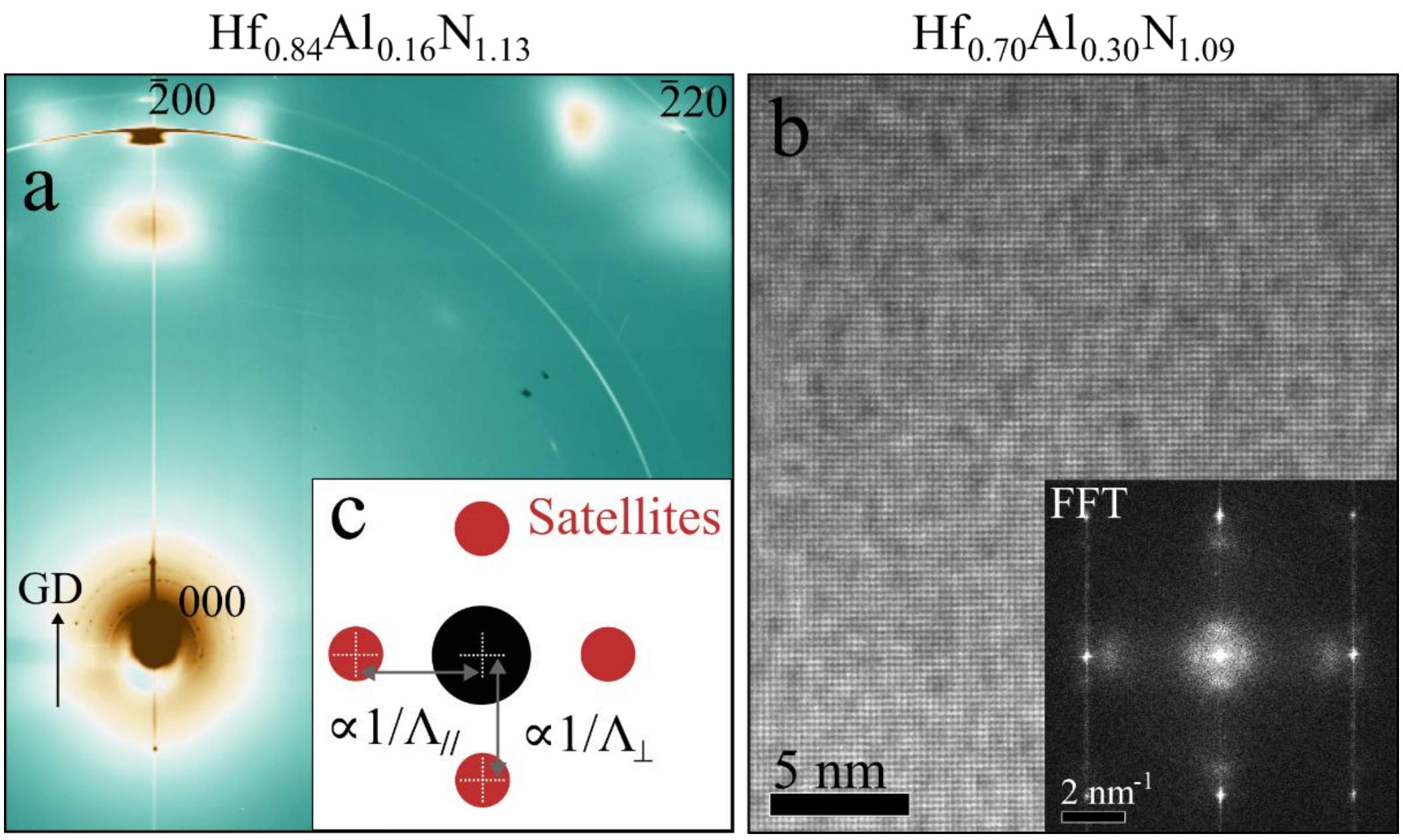

*Figure 4: a) Synchrotron WAXS of $Hf_{0.84}Al_{0.16}N_{1.13}$ with clear satellite peaks around the main peaks. A slight sample misalignment results in weak intensity main peaks. b) Cross-sectional HAADF STEM of c-$Hf_{0.70}Al_{0.30}N_{1.09}$ in the [001] zone axis, with an inset FFT showing the satellites of the compositionally modulated superstructure. c) Illustration of a main peak and satellite peaks (in a cross-section), where the satellite distance corresponds to the period length in the respective direction.*

The satellite to main peak spacing is inversely proportional to the superstructure period in that particular direction, as illustrated in Figure 4c, enabling straightforward comparison of the domain sizes in the lateral ($\Lambda_{\parallel}$) and the transverse ($\Lambda_{\perp} = 1/k_{\perp}$) directions. By comparing the distances to the satellite in Figure 3d-f and Figure 4a, we can determine that the checkerboard superstructure is composed of nominally equally sized cubelet domains, in all three perpendicular directions: [100], [010], and [001]. Considering a large measurement uncertainty in localizing the center of the wide, diffuse, and low-intensity satellite peaks, the periodicities calculated using SAED and WAXS data show small differences between the transverse and lateral directions, i.e., for $Hf_{0.84}Al_{0.16}N_{1.13}$ and $Hf_{0.70}Al_{0.30}N_{1.09}$ the transverse period is about 1-1.5 Å larger than the lateral one. The similarity between $\Lambda_{\perp}$ and $\Lambda_{\parallel}$ indicates that the c-HfAlN can minimize heterogenous strain in the lattice if the checkerboard is equal in the three <100> directions.

The transverse period can also be calculated from the XRD peak position from Figure 1 using Equation *3.1* [49], [50], where $\theta_0$ is the angle of the main peak, $\theta_n$ is the angle of the satellite, and n (= 1) is the order of the satellite. As shown in Figure 5, plotting the period against Al-content reveals a nearly linear trend for films in Series B, with higher Al concentration corresponding to larger domain sizes and the values of $\Lambda_{\perp}$ agrees well between the techniques. The larger period of Series A is explained by more energetic deposition conditions, see section 3.4.1. The smallest period of approximately 7.5 Å is observed in overstoichiometric $HfN_{1.22}$ and $HfN_{1.33}$ (due to self-organized point defects [40]), and the largest period of about 13.1 Å in $Hf_{0.71}Al_{0.29}N_{1.15}$ and $Hf_{0.70}Al_{0.30}N_{1.09}$. The large FWHM of the satellite peaks suggests significant variations in the superstructure period, allowing only an approximate average to be estimated using the center of the satellite peak at max intensity.

$$\Lambda = \frac{n\lambda}{2|\sin\theta_0 - \sin\theta_n|} \tag{3.1}$$

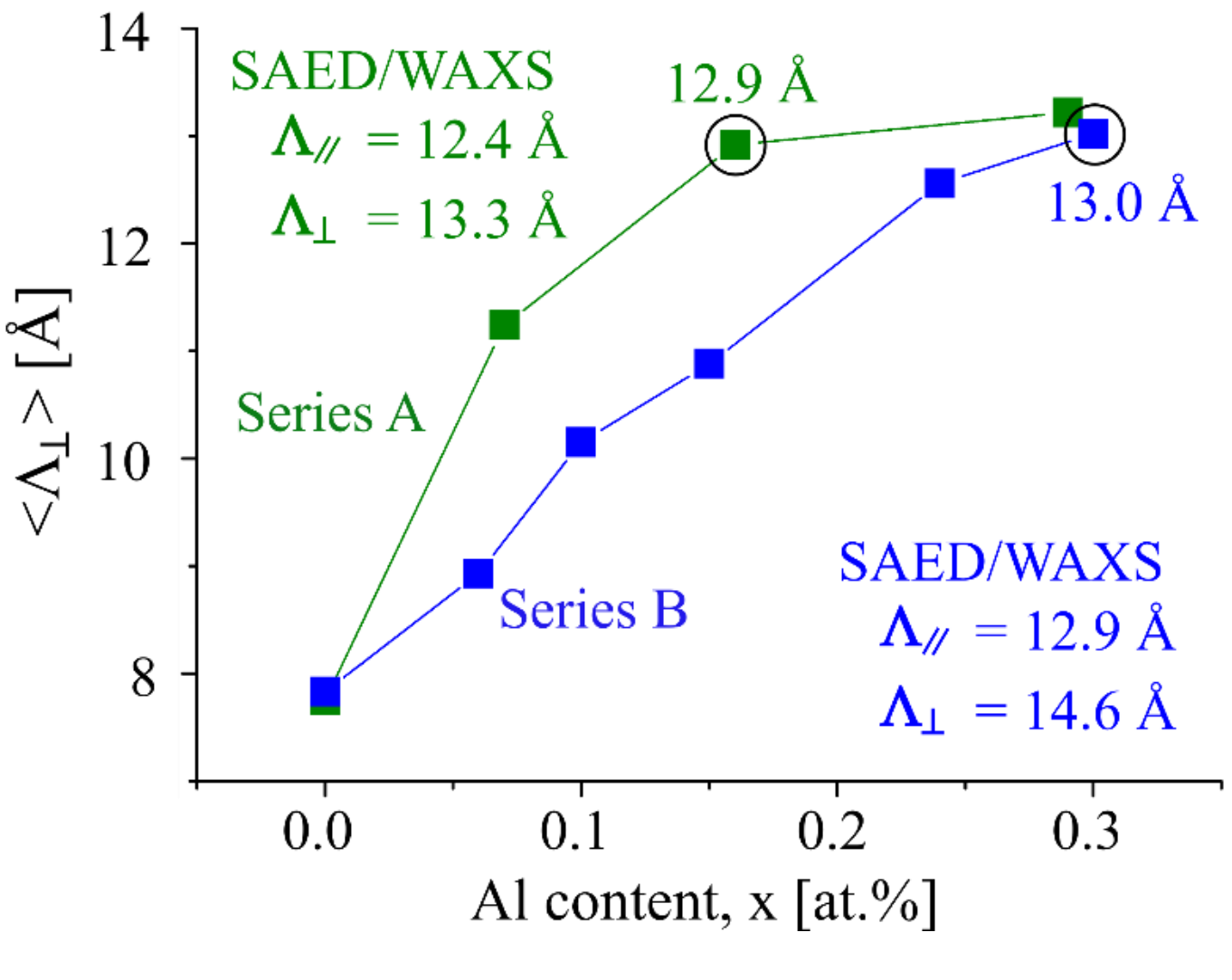

*Figure 5: Average transverse size of the superstructure as a function of Al concentration, x, in c-$Hf_{1-x}Al_xN_y$, calculated from XRD data in Figure 1. The additional numbers are values obtained by SAED and WAXS for $Hf_{0.84}Al_{0.16}N_{1.13}$ (green) and $Hf_{0.70}Al_{0.30}N_{1.09}$ (blue).*

### 3.4.1. Effect of Deposition Parameters on Crystal Quality and Superstructure Formation

Several deposition parameters influence the crystalline quality and the superstructure formation in epitaxially grown HfAlN on MgO(001) substrates. We have investigated three key parameters: a) growth temperature, b) ion assistance, and c) the often-neglected backscattered Ar neutrals.

#### *3.1.1.1. Effect of Growth Temperature*

The films in Series A and Series B were all grown at high temperatures, 800 °C, which is sufficient to initiate bulk spinodal decomposition. To investigate the effects of substrate temperature, $Hf_{0.76}Al_{0.24}N_{1.15}$ from Series B were re-deposited at three different substrate temperatures, but otherwise identical conditions: 300 °C, 600 °C, and 900 °C. The XRD results in Figure 6a show that the high-quality single-crystal structure was maintained independent of deposition temperature, and with a near identical superstructure period of ~12.5 Å. A higher substrate temperature appears to lower the superstructure quality, as the satellite peak width is substantially smaller for 300 °C compared to 900 °C. On the other hand, at 300 °C, the film peaks and satellites have all shifted towards smaller angles by $\Delta 2\theta^{002} = 0.45°$ and $\Delta 2\theta^{004} = 1.17°$ compared to the films grown at 600°C and 900°C, which indicates presence of a large compressive stress. The large compressive stress is more clear for the 004 peak, see supplementary information section 5. The high crystalline quality and a well-defined superstructure satellite peak even at 300 °C, where typically a fully random solid solution would be expected, indicates that substrate temperature plays only a minor role for the film structure.

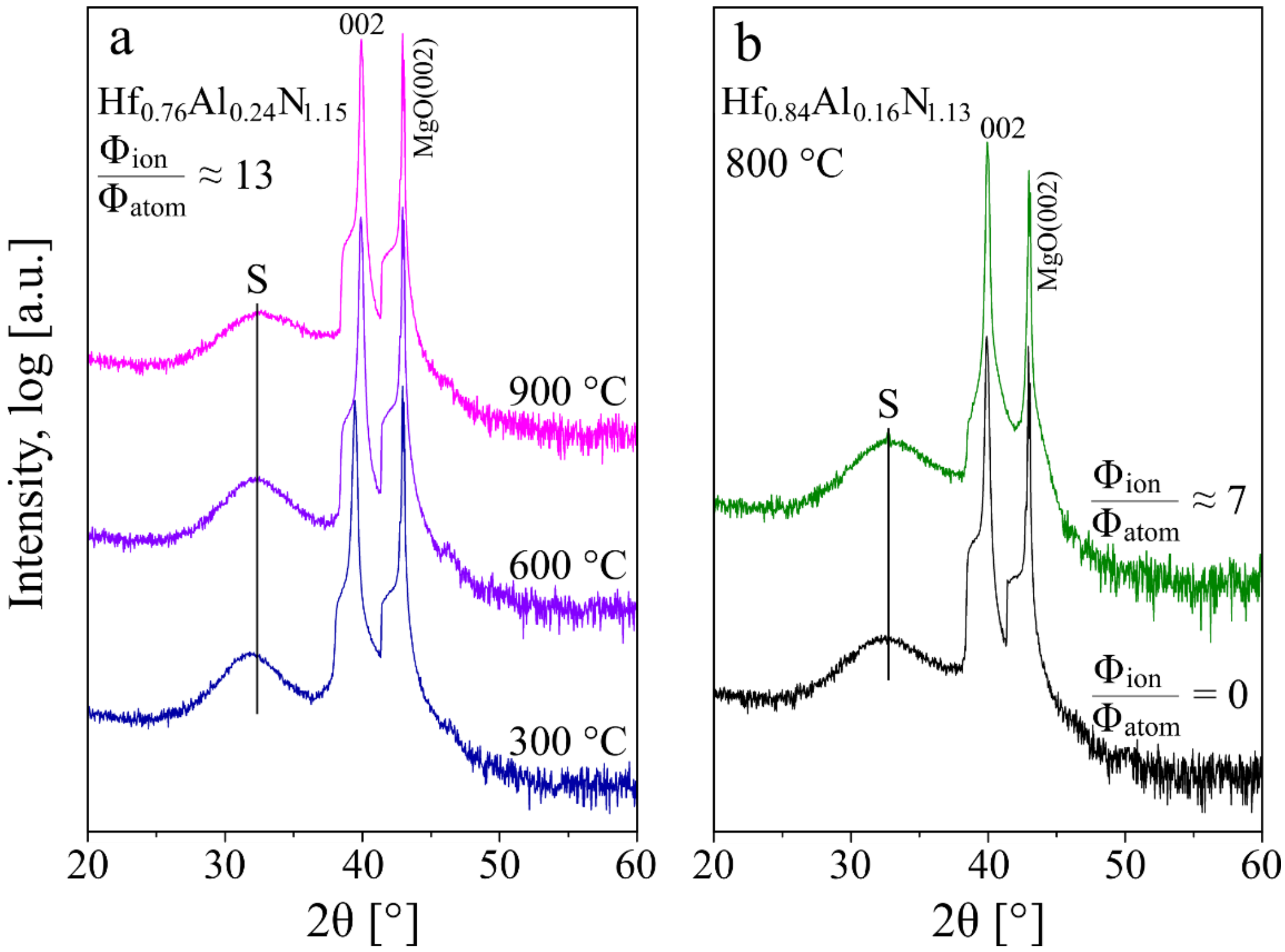

*Figure 6: Selected range of XRD $\theta$–$2\theta$ scans from a) c-$Hf_{0.76}Al_{0.24}N_{1.15}$ (Series B) grown at different temperatures, but otherwise identical conditions including ion assistance with an ion energy of 20 eV and a ion-to-atom flux ratio of 13, and b) c-$Hf_{0.86}Al_{0.16}N_{1.13}$ (Series A) grown at 800 °C with and without ion assistance, but otherwise identical conditions.*

#### 3.1.1.2. *Effect of Ion Assistance*

Each deposited atom in Series A and B is accompanied by a large number of bombarding ions of relatively low energy, see Table 1, extracted from the plasma. In Series B, the ion-to-atom flux ratio was fairly stable between 13 to 15, and each ion had an energy of about 20 eV. This large amount of energy (more than 260 eV per atom) incident on the growing film consequently changes the growth conditions and thus may affect the film structure [51]. It was shown for c-$Ti_{0.5}Al_{0.5}N$ grown at ~250°C, that a substantially improved crystalline quality and high density was obtained when the film was grown using an ion energy of about 20 eV, at a sufficiently high flux [27]. Low energy ions, below approximately 30 eV, do not have enough energy to displace film atoms that have already been bonded at the lattice sites, neither bulk nor surface. Instead, the quality improvements were attributed to an increased mobility of adsorbed atoms and small clusters. The effect on growth of HfAlN was therefore investigated by re-depositing the films in Series A while turning off ion assistance, see the comparison of $Hf_{0.84}Al_{0.16}N_{1.13}$ in Figure 6b. Surprisingly, the film grown without ion assistance exhibit a similar high crystalline quality and also a very similar superstructure period. Thus, like the substrate temperature, the use of ion assistance has only minor effects on the film formation which can be explained by another source of energetic particles, namely backscattered Ar atoms.

#### 3.1.1.3. *Effect of Backscattered Ar Neutrals*

Neutral backscattered Ar atoms are particularly important to consider when sputtering heavy elements using a light working gas, as is the case with Hf (178 amu) sputtering in an Ar (40 amu) atmosphere. Due to collision kinematics, the probability of backscattering as well as the energy of the reflected neutrals increases with increasing mass difference. The energy and flux of backscattered particles are quite challenging to measure experimentally. Instead, we have performed simulations on the sputtering process using SDTrimSP [52], [53] for reactive sputtering of both Hf and Al targets, and particle transport through the sputtering gas using SIMTRA [54], [55]. The targets were assumed to be fully poisoned with a surface composed of HfN and AlN, respectively, since the $N_2$ partial pressure is relatively high. More details on the simulations can be found in supplementary information section 6.

Figure 7a shows the energy distribution of the Ar neutrals just after being backscattered at the targets. The applied voltage to the targets (-340 V to -400 V) defines the initial Ar energy and scales the resulting energy distribution. The Ar energy from the AlN target at 400 eV is low, as expected from the lighter mass of Al (27 amu) and N (14 amu) compared to Ar. The backscatter yield, β, for 400 eV Ar is calculated to $\beta_{\mathrm{HfN}} = 2.22 \times 10^{-1}$, and $\beta_{\mathrm{AlN}} = 3.63 \times 10^{-4}$. Given the three orders of magnitude lower $\beta_{\mathrm{AlN}}$, and the low energy, we focus mainly on backscattered Ar from the Hf target. For the HfN target, a broad energy distribution of backscattered Ar is observed, with a significant fraction reaching up to 250 eV. The small peaks in Figure 7a near 150 eV likely results from binary collisions between Ar and Hf atoms at the target surface. The maximal energy transfer, $T_{max}$, assuming a head-on elastic binary collision is given by Equation *3.2* where $M_1$ and $M_2$ are the mass of the ion and target atom respectively. The term before $E_{ion}$, called the kinematic factor, determines the maximal possible energy transfer that may occur in the collision.

$$T_{max} = \frac{4M_1M_2}{(M_1 + M_2)^2} E_{ion}. \quad (3.2)$$

After the backscatter event, the Ar atom travels through the gas towards the substrate, undergoing collisions along the way. This transport was simulated in SIMTRA using the initial energy and angular distributions provided by SDTrimSP and the chamber geometries, notably the 12.5 cm target-to-substrate distance. Figure 7b shows the resulting energy distributions at the substrate, where gas phase collisions decreased the Ar energy significantly. A target voltage of 400 V results in an average energy of about 50.3 eV at the substrate. However, due to the skewness of the distribution, a large fraction of the Ar arrives with a substantially larger energy, e.g., 31.5 % arrive with an energy above 70 eV.

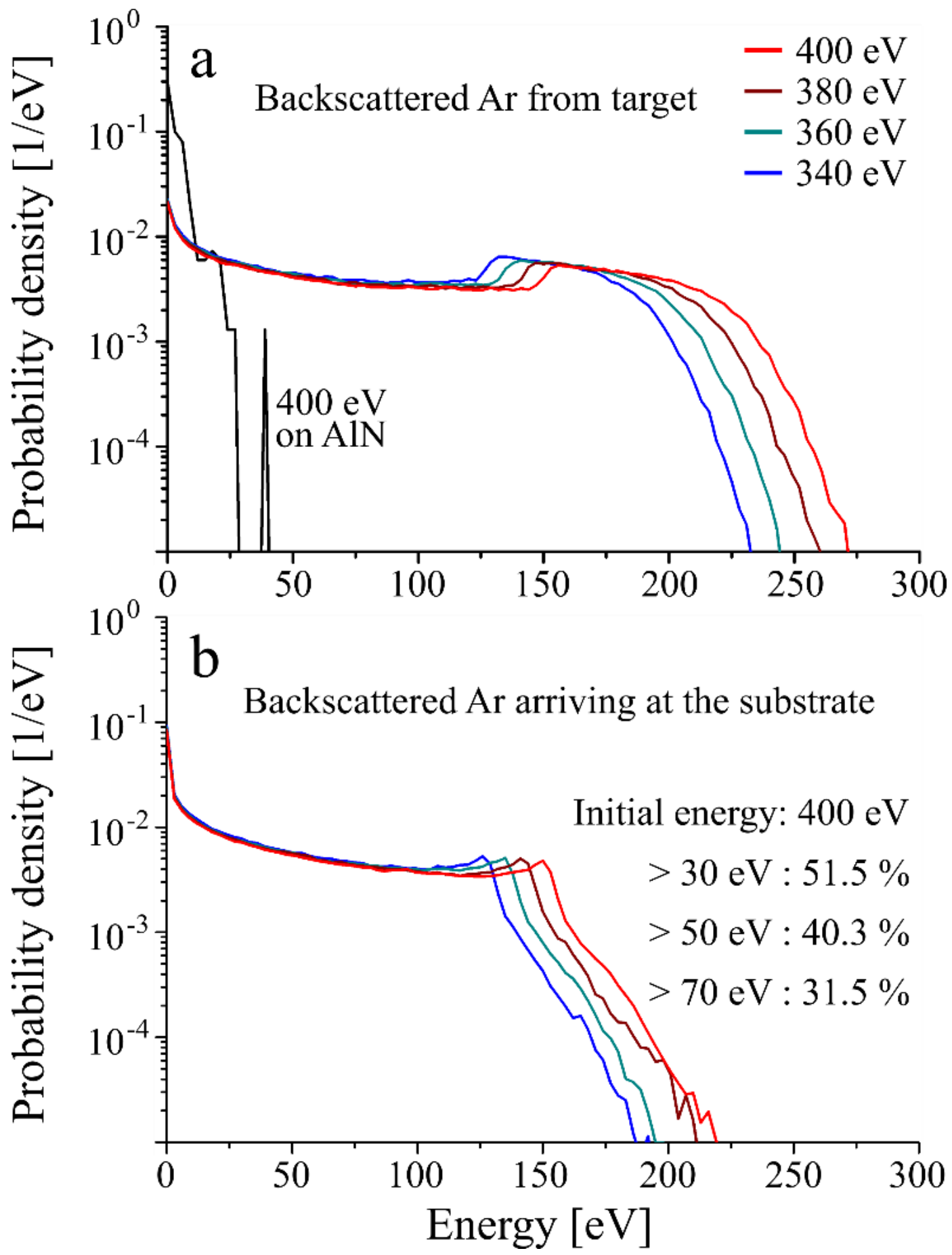

*Figure 7: Energy distributions of backscattered Ar atoms as a function of the applied target voltage. a) At reflection from the HfN target surface, b) as the Ar atoms arrive at the substrate. The numbers in b) show the percentage of particles arriving at the substrate with an energy higher than 30 eV, 50 eV and 70 eV, illustrating the skewness of the distribution at 400 V target voltage.*

The Hf and Al sputter yield and energies from the poisoned targets was calculated using the model of Depla and Van Bever [56], with details in supplementary information section 6. The average energy of both Hf and Al was below 10 eV on arrival to the substrate, which is too low for a significant effect on the film formation, but will slightly improve the adatom mobility. Using the Hf sputter yield, the flux ratio between arriving Ar neutrals and film forming atoms was calculated according to Table 2, with input from the deposition parameters, density and chemical composition of respective films. The relatively high flux of backscattered neutrals for HfAlN, ratio of ~0.45 for Series A, and ~0.7 for Series B (see Table 2), combined with the wide energy distribution, can significantly influence the film structure. For example, if their energy exceeds the threshold displacement energy, $E_d$, they can displace a lattice atom, forming a vacancy-interstitial (Frenkel) pair, where $E_d$ is both material and crystallographic-direction dependent [57]. Based on the higher melting point of HfN, the displacement energy for HfN and HfAlN is estimated to be equal or higher than the available values for ZrN, reported as a weighted average to $E_d^{Zr}$ = 33 eV and $E_d^{N}$ = 29 eV [58]. A large fraction of the backscattered neutrals has higher energy than $E_d$, see Figure 7b, making atomic displacements likely, potentially explaining the high-quality films and well-defined superstructure formation. In addition, the relatively high Ar content (~1-2 at. %, see Table 1), suggests that energetic Ar may become trapped in the lattice. Note that surface displacements may also occur, where the energy threshold, $E_d^{(s)}$, is often approximated to half of $E_d$.

*Table 2: Hf target voltage during deposition. The average kinematic factor determines the average transferred energy from an incident ion to film atom using equation* 3.2*. Two versions of the flux ratio between backscattered Ar and film atom are given: one for the total Ar flux, the other for the Ar flux with an energy above 43 eV.*

| | Hf Target voltage [V] | Average kinematic factor | Total backscattered Ar/atom flux ratio | Backscattered Ar/atom flux ratio (Ar > 43 eV) |
|---|---|---|---|---|
| Series A | | | | |
| $HfN_{1.22}$ | 412 | 0.692 | 0.36 | 0.16 |
| $Hf_{0.93}Al_{0.07}N_{1.15}$ | 383 | 0.701 | 0.46 | 0.20 |
| $Hf_{0.84}Al_{0.16}N_{1.13}$ | 367 | 0.716 | 0.49 | 0.20 |
| $Hf_{0.71}Al_{0.29}N_{1.15}$ | 360 | 0.738 | 0.46 | 0.19 |
| $Hf_{0.59}Al_{0.41}N_{1.23}$ | 353 | 0.759 | 0.47 | 0.19 |
| $Hf_{0.29}Al_{0.71}N_{1.04}$ | 339 | 0.812 | - | - |
| Series B | | | | |
| $HfN_{1.33}$ | 351 | 0.695 | 0.74 | 0.30 |
| $Hf_{0.94}Al_{0.06}N_{1.17}$ | 350 | 0.700 | 0.81 | 0.32 |
| $Hf_{0.90}Al_{0.10}N_{1.12}$ | 349 | 0.705 | 0.77 | 0.31 |
| $Hf_{0.85}Al_{0.15}N_{1.11}$ | 350 | 0.714 | 0.72 | 0.29 |
| $Hf_{0.76}Al_{0.24}N_{1.15}$ | 351 | 0.730 | 0.69 | 0.27 |
| $Hf_{0.70}Al_{0.30}N_{1.09}$ | 352 | 0.739 | 0.61 | 0.24 |

Brice et al. proposed a model linking incident particle energy (ion or atom) to displacement effects in the growing film [59] as the particle continuously deposits energy while traversing through the material. Three different energy regions were identified: In region I, the particle energy is too low to cause any displacements. In region II, the energy is enough for surface displacements, but low enough to avoid bulk displacements. In region III, the particle has enough energy for bulk displacements. The specific value of the threshold energies between the regions depends on the surface and bulk displacement energy of the particular material, as well as the masses of the involved atoms, i.e. the energy transfer efficiency, why displacement of Al may be more effective than Hf for energetic Ar bombardment. Although we do not apply Brice´s model quantitatively to the HfAlN system due to unknown material parameters, it provides a useful conceptual framework. It has been successfully employed and experimentally validated in simpler systems such as metallic Ni/V multilayers [28]. Moreover, the model can explain the results for $TiB_2/CrB_2$ superlattices [60], where ion energies in Region II and Region III led to significant improvements in the superlattice quality.

Both model and experiments show that a significantly larger incident particle energy is required compared to $E_d$ to initiate displacements. We have estimated that an Ar energy higher than ~40-50 eV is required for region III in the case of HfAlN. The wide energy distributions of backscattered Ar neutrals thus span all three regions, where ~40 % of the neutrals have an energy higher than 50 eV and therefore fall within region III. Using Equation *3.2*, the minimum energy transfer necessary to overcome $E_d$ was calculated to 43 eV and the flux ratio when considering only Ar neutrals above this energy is shown in Table 2. Flux ratios of ~0.2 (Series A) and ~0.3 (Series B) indicate that for every 5-3 film forming atoms, one Ar arrives with sufficient energy to cause at least one bulk displacement. Note that the high flux of low energy (<25 eV) ion assistance falls within region I, causing no surface or bulk displacements.

With these results, we attribute the high crystalline quality of the cubic HfAlN thin films to backscattered Ar neutrals: Region I neutrals enhance adatom mobility and region II neutrals cause surface displacements that improve ordering and reduce grain boundary porosity, and region III neutrals induce sub-surface defects and densification. Moreover, we hypothesize that the 3D checkerboard superstructure is triggered by the high energy Ar neutrals via spinodal decomposition just below the surface of the growing film as the bulk displacements disturb the lattice and input energy that may be enough to enable bulk diffusion. The hypothesis is supported by the superstructure formation even at low temperatures where thermal decomposition is unlikely, see Figure 6. In addition, the larger superstructure period of Series A compared to Series B (c.f. Figure 5) may originate from the higher energy of the backscattered Ar for Series A, since the energy distribution scales with the applied target voltage, see Table 2. Similarly, the linear period with respect to Al concentration of Series B may be attributed to the stable energy distribution and flux of the backscattered Ar.

It is noteworthy that backscattered Ar neutrals are intrinsic to sputtering heavy elements and are difficult to control. Their energy can be reduced via increased gas pressure, although not enough in our setup to fully suppress region III energies. A definitive experiment on the role of backscattered Ar could involve switching out Ar for a heavier working gas (e.g., Kr or Xe) to eliminate backscattering or shift energy transfer dynamics, but such experiments are beyond the scope of this study.

#### 3.4.2. Formation of 3D Superstructure

To our knowledge, the present case of formation of a 3D superstructure from displacement initiated spinodal decomposition during film growth has not been reported before. The closest phenomena is 3D spinodal decomposition of TiAlN films during post-deposition annealing at temperatures high enough to activate bulk diffusivity [7]. Nevertheless, we believe that this structure could be possible to find in growth of other metastable TM-Al-N material systems under similar deposition conditions, albeit detection may be more demanding depending on the contrast between the domains. For context, exclusive (traditional) surface-initiated spinodal decomposition has previously been reported for epitaxial $Ti_{0.5}Al_{0.5}N$ [8]. However, the decomposition occurred only within a narrow deposition temperature range of 540-560 °C, forming long and thin coherent isostructural rectangular columns in the growth direction. In contrast, HfAlN forms smaller 3D cubelets that are stable within a wide temperature range of at least 300-900 °C. In $Zr_{0.64}Al_{0.36}N$, a random solid solution is obtained up until 700 °C where decomposition leads to phase-separated lamellar nanolabyrinthine structures, fully evolved at 900 °C [14], [61], rather than isostructural cubic domains. A high flux of low energy ion assistance was applied for both TiAlN and ZrAlN, which may explain the good crystalline quality, however the number of backscattered Ar can be neglected, as expected due to the lower mass of Ti and Zr.

These three different decomposition behaviors among group IV TM-Al-N material systems highlight the complex interplay of chemical driving force, strain energy, bonding and volume expansion of resulting phases versus parent phase [11], [12], [13], as well as the surface conditions during growth. While mixing enthalpy and strain energy help explain the delayed wurtzite formation in HfAlN compared to TiAlN [11], they do not explain the vastly different behavior between HfAlN and ZrAlN as the properties are very similar. This suggests that additional factors play a critical role, where we propose that the energetic backscattered Ar during growth modifies the surface conditions to promote formation of coherent cubical domains in c-HfAlN but not ZrAlN.

### 3.5. Structure at high Al content: Nanocrystalline Wurtzite h-HfAlN

Increasing the Al content to x > 0.41 results in a phase transition from cubic to wurtzite crystal structure of the films. Figure 8 shows HAADF STEM images of h-$Hf_{0.59}Al_{0.41}N_{1.23}$, exhibiting a highly distorted wurtzite lattice with a preferred (0001) texture along the surface normal, in agreement with the XRD results in Figure 1. An epitaxially stabilized cubic phase is observed near the MgO substrate up to a thickness of about 40 nm, beyond which the film relaxes into the more stable wurtzite structure. As discussed above, the incorporation of large Hf atoms significantly expands the wurtzite lattice, resulting in a pronounced shift in the XRD peak towards lower 2θ angles. This lattice distortion also degrades the crystalline quality, evident from the broad XRD peak and the wavelike lattice pattern in Figure 8b with no visible grain boundaries. The h-HfAlN film exhibits decomposition into Hf- and Al-rich nitride nanodomains, visible as bright and dark contrast, respectively. Unlike the cubic superstructure, these domains are more elongated laterally, measured approximately 2-3 times longer than in the transverse direction. The SAED inset in Figure 8a confirms epitaxial growth of c-HfAlN near the MgO interface, while the wurtzite phase displays a 0001-fiber texture, in agreement with XRD results. Although no satellite peaks were observed, likely due to the poor crystalline quality, some degree of ordering is evident in Figure 8b, where alternating Hf-rich and Al-rich domains appear. Fiber-textured nanocrystalline structures are commonly observed in wurtzite transition-metal aluminum nitrides such as h-TiAlN [62], [63] and h-ZrAlN [14], [64], thus h-HfAlN thin films develop similarly to the aforementioned material systems.

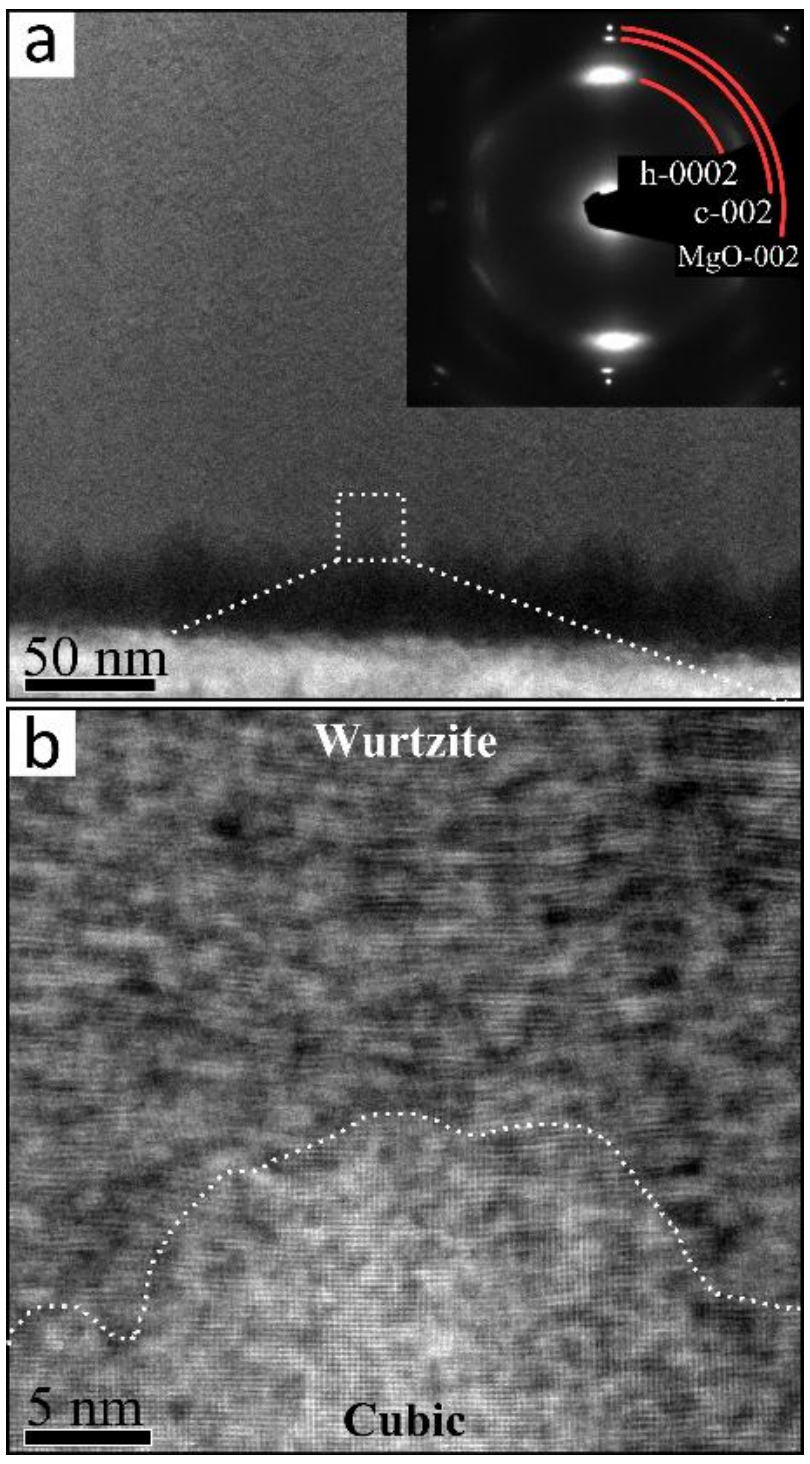

*Figure 8: Cross-sectional a) overview and b) HR HAADF STEM micrographs of h-$Hf_{0.59}Al_{0.41}N_{1.23}$ film. The inset in a) shows a SAED pattern from the film, and MgO substrate, including the thin epitaxially stabilized cubic $Hf_{0.59}Al_{0.41}N_{1.23}$ phase. The dotted line in b) marks the boundary between cubic and wurtzite-hexagonal phase.*

### 3.6. Mechanical Properties of HfAlN Thin Films

The hardness, yield stress and plasticity of the $Hf_{1-x}Al_xN_y$ thin films were evaluated using nanoindentation and micropillar compression testing of cubic and hexagonal phase films.

#### 3.6.1. Nanoindentation Hardness

Figure 9a shows that by alloying HfN with Al significantly increases the hardness of the cubic phase from approximately 26 GPa to 36-41 GPa, which then remains fairly stable. At x = 0.41, the lattice transforms to the softer wurtzite phase, with a hardness drop to about 22 GPa, and remains more or less constant up to x = 0.71. The comparatively high hardness of h-HfAlN relative to h-AlN (H = 11 GPa under similar deposition conditions [14]) is likely due to Hall-Petch hardening arising from the nanocrystalline grain structure.

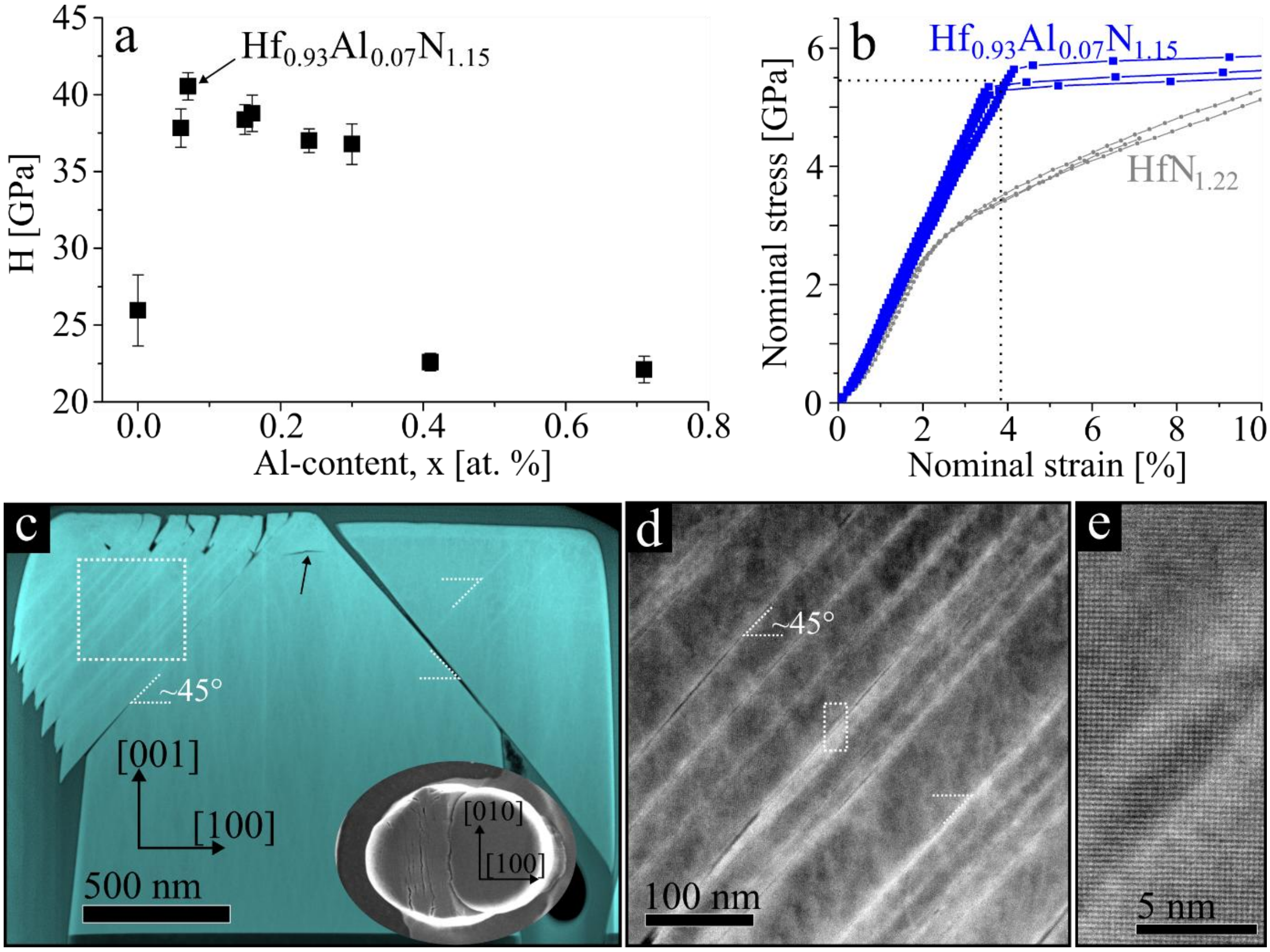

*Figure 9: a) Hardness of $Hf_{1-x}Al_xN$ as a function of Al content, x. b) Engineering stress-strain curves of c-$Hf_{0.93}Al_{0.07}N_{1.15}$ in micropillar compression, and data for c-$HfN_{1.22}$ acquired from similar pillar geometry and indenter is presented for comparison, reproduced with permission [40]. c-e) HAADF STEM micrographs of a lamella extracted from a compressed pillar: c) colored overview image that highlights the dislocations and fractures, and an inset of a top-view SEM micrograph where the axis shows the crystallographic directions. The dotted wedge markings indicate 45° angle. d-e) HR HAADF STEM of the fractured area marked in c) and d) respectively.*

### 3.6.2. Yield Strength and Fracture in cubic HfAlN

The stress-strain curves from micropillar compression tests of c-$Hf_{0.93}Al_{0.07}N_{1.15}$, exhibit a linear elastic deformation up to nearly 4 % strain, reaching a high stress of approximately 5.5 GPa before strain-burst occurs and the pillars fractures, as shown in Figure 9b . This behavior stands in substantial contrast to that of Al free samples c-$HfN_{1.22}$ and c-$HfN_{1.33}$ from series A and Series B, whose deformation behavior was previously reported [40]. In their deformation, significant ductility and strain hardening was observed as the material plastically deformed in a metal-like fashion via dislocation motion on the {111}<110> slip system [40]. Here, the addition of just 7 at.% Al on the metal sublattice results in nearly doubled the elastic strain and doubled nominal stress at the yield point but also shows brittle fracture. The fracture behavior of c-$Hf_{0.93}Al_{0.07}N_{1.15}$ was further examined using HAADF STEM, following the extraction of an electron transparent lamella from a compressed micropillar via FIB. Figure 9c-e reveals massive slip and/or fracture occurring along planes oriented approximately 45° to the surface normal, with all fractures confined to the upper half of the pillar. Based on the orientation of the fractured planes and directions, the active slip system is identified as {110}<110>. No additional dislocations are observed in the compressed pillar, compared to the as-deposited film. Figure 9d-e show indication of both fractures and extensive slip. In some areas, material free gaps are visible, while in others, lattice continuity is observed across the crack.

The improved hardness and stress strain behavior of c-HfAlN suggests that nucleation of dislocations is unfavorable, and the numerous threading dislocations in the as-grown film remain immobile (see supplementary information section 7). This is attributed to local strain fields arising from the compositionally modulated 3D checkerboard superstructure, which effectively pins the dislocations, similar to observations in spinodally decomposed c-TiAlN [9], [48]. When the built-up stress surpasses the yield strength, the material fails through sudden fracture rather than dislocation enabled plastic deformation. The immobilized dislocations may explain the switch of activated slip system to {110}<110> since the largest resolved shear stress (Schmid factor) in this experimental setup is found for {110} planes, rather than {111}. Moreover, the fact that the chemical composition affects slip system activation is well established in literature, for instance: understoichiometric c-$ZrN_y$ favors {111}<110>, while near-stoichiometric c-ZrN prefers {110}<110> [65], and incorporation of Al into c-TiAlN, results in a shift to {111}<110> slip system, in contrast to the well-known {110}<110> observed in c-TiN [66], indicating the opposite trend compared to c-Hf(Al)N. In the present case, vertical cracks formed at the top surface of the micropillar, visible in Figure 9c, are likely caused by frictional forces between the diamond tip and film surface. These cracks were subsequently deflected by 45° on to the {110}<110> slip system, about 100 nm beneath the surface. This behavior is rather unusual for nominally brittle ceramics, which typically fail via catastrophic fracture, as seen in materials like c-Cr(Al)N [67], spinel-cubic $MgAl_2O_4$ [68], c-TiN [69], c-Ti-C-N and c-Zr-C-N [70], and c-WC [71].

### 3.6.3. Plasticity of wurtzite HfAlN

Micropillar compression was performed on ~400 nm diameter pillars of h-$Hf_{0.59}Al_{0.41}N_{1.23}$ to evaluate the mechanical response of nanocrystalline wurtzite phase film. Figure 10a shows the resulting engineering stress-strain curves and Figure 10b-e displays SEM micrographs of the as-milled (b) and post-mortem (c-e) pillars. A high yield stress of 5.3 ± 0.3 GPa was obtained, followed by a limited plastic deformation, as indicated by a deviation from linearity in the stress-strain curves while stress continues to increase. Upon further loading, strain bursts occur as the pillars fracture in a stepwise manner shown by the halted burst at ~14 % strain. The loading was stopped just after the first strain burst for the pillar in Figure 10c. Multiple fractures can be seen located near the top where the stress is highest due to pillar taper [72]. Continuing the loading past this point results in additional fractures in the middle and bottom of the pillars, and eventually catastrophic failure, as shown in Figure 10d-e. Fractures formed at approximately 45° to the pillar axis, consistent with the direction of maximum resolved shear stress. No preferred in-plane fracture direction was observed, with fracture surfaces appearing randomly oriented which reflects the fiber textured nanocrystalline nature of the film.

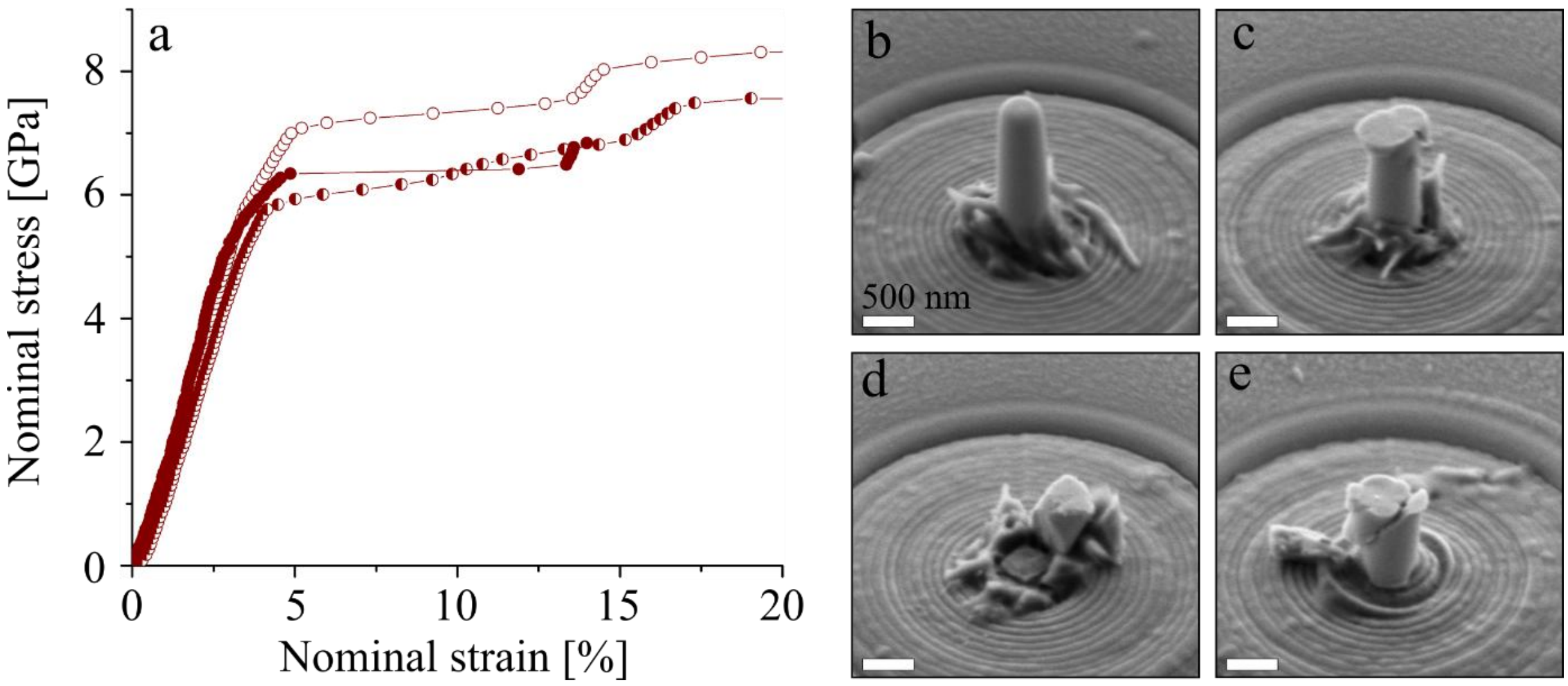

*Figure 10: Micropillar compression results from h-$Hf_{0.59}Al_{0.41}N_{1.23}$. a) Engineering stress-strain curves. Representative SEM micrographs of b) as-milled pillar, and c-e) compressed pillars.*

The initial plasticity and the plasticity just after a strain burst may stem from grain boundary sliding and reorganization of the nanocrystalline microstructure, since dislocation mediated plasticity is not feasible in very small grains. Similar properties were seen in compression of nanocrystalline CrAlSiN [73], however, the microstructure with nanosized CrAlN grains within an amorphous $Si_3N_4$ matrix allowed a larger plasticity compared to the partly decomposed nanocrystalline h-HfAlN in this work.

The yield stress of h-$Hf_{0.59}Al_{0.41}N_{1.23}$ (~5.3 GPa) is surprisingly similar to that of the c-$Hf_{0.93}Al_{0.07}N_{1.15}$ film (~5.5 GPa), and the stress just before the first strain-burst is even higher (~6.3 GPa), partly attributed to Hall-Petch strengthening from the nanocrystalline structure. However, differences in pillar geometry and size must be considered, as these can strongly influence the measured stress–strain response [72]. Smaller pillar diameters were required for h-$Hf_{0.59}Al_{0.41}N_{1.23}$, due to thinner film thickness, and the well-known size effect [74] likely leads to an overestimation of the strength. A more direct comparison can be made with c-$HfN_{1.22}$ from previous work [40], which had similar pillar dimensions and a lower a yield stress (~4.5 GPa). Thus, h-$Hf_{0.59}Al_{0.41}N_{1.23}$ exhibits higher strength than c-$HfN_{1.22}$. As previously shown for $HfN_{1.22}$, larger pillar diameter and taper reduces the measured yield stress, suggesting the strength of c-$Hf_{0.93}Al_{0.07}N_{1.15}$ may be underestimated in Figure 9b is and the strength of h-$Hf_{0.59}Al_{0.41}N_{1.23}$ in Figure 10a may be overestimated. Still, h-HfAlN demonstrates promising behavior, combining high strength with the onset of plasticity prior to fracture.

# 4. Conclusions

Single-crystal rocksalt (c-$Hf_{1-x}Al_xN_y$) and nanocrystalline wurtzite (h-$Hf_{1-x}Al_xN_y$) coatings were synthesized by ion-assisted magnetron sputtering onto MgO(001) substrates and thoroughly characterized using ToF-ERDA, RBS, XRD, XPS, SEM, HR STEM, SAED, and synchrotron WAXS. Cubic films with Al content x < 0.30 exhibited a cube-on-cube epitaxy and a 3D checkerboard-like superstructure aligned along the <001> directions, as evident by satellites in diffraction analysis. This structure is attributed to sub-surface spinodal decomposition triggered by energetic backscattered Ar neutrals from the Hf target. Sputtering simulations from a poisoned HfN target surface indicate that a high energy Ar flux was obtained, which induces atomic (bulk) displacements that promote compositional modulations within a wide substrate temperature window (300 °C - 900 °C). Experiments show only minor effects of substrate temperature and ion assistance on the crystal quality and superstructure formation.

The superstructure period increased linearly with Al concentration from 7.5 Å for pure $HfN_y$ to 13 Å for x= 0.3, remaining sensitive to the energy distribution of the backscattered Ar neutrals. Higher Al contents (x > 0.41) yielded nanocrystalline wurtzite films with strong 0001 texture. While HR-STEM revealed phase separation into Al- and Hf-rich domains, no diffraction satellites were observed in this structure. XPS confirmed two distinct chemical environments for Hf and N in the cubic and wurtzite phases.

This work presents the first direct evidence of 3D superstructure formation during film growth via displacement-initiated spinodal decomposition. We propose that such structures may form in other metastable TM-Al-N systems under similar conditions, though their detection may be more challenging. The distinct decomposition behaviors observed across group IV TM-Al-N alloys underscore the critical role of surface conditions during growth. In particular, energetic backscattered Ar neutrals appear to promote coherent cubic domain formation in Hf-Al-N, highlighting a previously underappreciated mechanism for directing nanostructural evolution in complex nitrides.

Mechanical testing showed a substantial hardness in the cubic phase to ~36 GPa for x=0.06-0.3 due to strain fields from the 3D checkerboard superstructure. The nanocrystalline wurtzite phase exhibited hardness of ~22 GPa over a broad Al range (x=0.4-0.7), attributed to Hall-Petch hardening. Micropillar compression of c-$Hf_{0.93}Al_{0.07}N_{0.15}$ revealed no dislocation activity; instead, fracture occurred through strain burst at nearly double the stress and strain compared to $HfN_{1.22}$. The activated slip system {110}<110> differs from the typical {111}<110> in $HfN_y$, suggesting a modification due to alloying. Fracture paths quickly deflected onto this preferred slip system. In contrast, the wurtzite pillars displayed some plasticity and a high yield strength, attributed to grain boundary sliding, before fracturing along planes inclined ~45° to the surface normal.

## 5. CRediT Autor Statement

**Marcus Lorentzon**: Conceptualization, Investigation, Validation, Data Curation, Visualization, Formal analysis, Writing – Original Draft, Writing – Review & Editing. **Naoki Takata**: Investigation, Writing – Review & Editing. **Diederik Depla**: Methodology, Software, Investigation, Formal analysis, Writing – Review & Editing. **Tianqi Zhu**: Investigation, Formal analysis. **Grzegorz Greczynski**: Investigation, Writing – Review & Editing. **Rainer Hahn**: Investigation, Writing – Review & Editing. **Anton Zubayer**: Formal analysis. **Justinas Palisaitis**: Investigation. **Helmut Riedl**: Funding acquisition. **Dasom Kim**: Investigation, Formal analysis. **Lars Hultman**: Supervision, Funding acquisition, Writing – Review & Editing. **Jens Birch**: Supervision, Funding acquisition. **Naureen Ghafoor**: Conceptualization, Supervision, Investigation, Funding acquisition, Writing – Original Draft, Writing – Review & Editing

## 6. Conflict of Interest

The authors declare no conflict of interest.

## 7. Acknowledgement

We acknowledge DESY (Hamburg, Germany), a member of the Helmholtz Association HGF, for the provision of experimental facilities. Parts of this research were carried out at Petra III. Data was collected using beamline P07 operated/provided by DESY Photon Science. We would like to thank Dr. Sanjay Nayak for assistance during the experiments. Beamtime was allocated for proposal I-20220687 EC. The authors gratefully acknowledge the financial support from the Swedish Research Council, VR, 2018-05190_VR and the Swedish Government Strategic Research Area in Materials Science on Advanced Functional Materials (AFM) at Linköping University (Faculty Grant SFO Mat LiU No. 2009 00971). The authors thanks MIRAI 2.0 seed-funding, connecting Swedish and Japanese universities, for collaboration initiation between Linköping University and Nagoya university. Ion beam accelerator operation was supported by Swedish Research Council VR-RFI (Contract No. 2019-00191). Swedish Research Council and Swedish Foundation for Strategic Research are acknowledged for access to ARTEMI, the Swedish National Infrastructure in Advanced Electron Microscopy (2021-00171 and RIF21-0026). The financial support by the Austrian Federal Ministry for Digital and Economic Affairs, the National Foundation for Research, Technology and Development, and the Christian Doppler Research Association is gratefully acknowledged (Christian Doppler Laboratory, Surface Engineering of High- performance Components).